\documentclass{aa}
\usepackage[utf8]{inputenc}

\usepackage{lscape}
\usepackage{hyperref}
\usepackage{graphicx}
\usepackage[dvipsnames]{xcolor}
\usepackage{subfigure}
\usepackage{natbib}
\usepackage{geometry}
\usepackage{changepage}
\usepackage{subcaption}
\usepackage{amsmath}
\usepackage{gensymb}
\usepackage{txfonts}

\bibpunct{(}{)}{;}{a}{}{,} 

\hypersetup{colorlinks,linkcolor={blue},citecolor={blue},urlcolor={blue}}

\begin{document}

\title{Astrometry in crowded fields towards the Galactic Bulge}
\author{
Alonso Luna\inst{\ref{ESO},\ref{UNAB}}
\and Tommaso Marchetti \inst{\ref{ESO}}
\and Marina Rejkuba \inst{\ref{ESO}}
\and Dante Minniti \inst{\ref{UNAB},\ref{VO},\ref{UFSC}}
}
\institute{
European Southern Observatory, Karl-Schwarzschild-Straße 2, 85748 Garching, Germany \label{ESO} \email{alonso.luna@eso.org}
\and Instituto de Astrofísica, Facultad de Ciencias Exactas, Universidad Andrés Bello, Fernández Concha 700, Las Condes, Santiago, Chile \label{UNAB}
\and Vatican Observatory, V00120 Vatican City State, Italy \label{VO}
\and Departamento de Fisica, Universidade Federal de Santa Catarina, Trinidade 88040-900, Florianopolis, Brazil \label{UFSC}
}

\date{\today}

\abstract
{The astrometry towards the Galactic Bulge is hampered by high stellar crowding and patchy extinction. This effect is particularly severe for optical surveys such as the European Space Agency satellite \textit{Gaia}.}
{In this study, we assess the consistency of proper motion measurements between optical (\textit{Gaia} DR3) and near-infrared (VIRAC2) catalogues in comparison with proper motions measured with the Hubble Space Telescope (HST) observations in several crowded fields towards the Galactic Bulge and in Galactic globular clusters.}
{Assuming that the proper motion measurements are well characterised, the uncertainty-normalised proper motion differences between pairs of catalogues are expected to follow a normal distribution. A deviation from a normal distribution defines the inflation factor $r$. By multiplying the proper motion uncertainties with the appropriate inflation factor values, the \textit{Gaia} (VIRAC2) proper motion measurements are brought into a 1$\sigma$ agreement with the HST proper motions.}
{The inflation factor ($r$) has a dependence on stellar surface density and for the brightest stars in our sample ($G<18$), there is a strong dependence on $G$-band magnitude, corresponding to the most precise \textit{Gaia} DR3 proper motions. In our study, we use the number of observed \textit{Gaia} DR3 sources as a proxy for stellar surface density. Assuming that the HST proper motion measurements are well determined and free from systematic errors, we find that \textit{Gaia} DR3 proper motion uncertainties are better characterised, having $r<1.5$, in fields with stellar number density under 200 \textit{Gaia} DR3 sources per arcmin$^{2}$, and are underestimated by up to a factor of 4 in fields with stellar densities larger than 300 sources per arcmin$^{2}$. For the most crowded fields in VIRAC2, the proper motion uncertainties are underestimated by a factor of 1.1 up to 1.5, with a dependence on $J$-band magnitude. In all fields, the brighter sources have the larger $r$ value.
At the faint end ($G$>19), the inflation factor is close to 1, meaning that the proper motions are already fully in agreement with the HST measurements within $1\sigma$.}
{ In the crowded fields that have both catalogues in common, VIRAC2 proper motions are in agreement with HST proper motions and do not need an inflation factor for their uncertainties. Given the depth and completeness of VIRAC2 in such fields, it is an ideal complement to \textit{Gaia} DR3 for proper motion studies towards the Galactic Bulge.}


\maketitle

\section{Introduction}
 
The Galactic Bulge\footnote{Throughout this work, the Bulge refers to the $300\,\deg^2$ region in $-10\deg<l<10\deg$ and $-10\deg<b<5\deg$, comprising a radius of $\sim$2.5 kpc around the Galactic Centre.} is the only inner region of a large galaxy where individual stars can be resolved from the ground. 
Therefore, the Galactic Bulge provides a possibility to study stellar interactions 
in high-density environments on galactic scales. As galaxies form inside-out, the studies of the stellar populations in the Galactic Bulge provide links to the early formation of the Milky Way \citep[e.g.][]{Barbuy2018, Zoccali2019, Fragkoudi2020}.
However, the highly variable reddening and stellar crowding make its characterisation a difficult task. Many studies thus focused on specific low reddening windows in the Bulge \citep[e.g.][]{Zoccali03, Clarkson_BTS2008, Johnson2011_PlautWindow, Bernard2018} or areas of particular interest, such as the Galactic Centre with the supermassive black hole \citep{gillessen2009, GRAVITY2018, GRAVITY2020} and the nuclear star cluster \citep[e.g.,][]{pfuhl2011, schoedel2014, FeldmeierKrause2015, Chatzopoulos2015, Nogueras-Lara2020}. Surveys of the inner Milky Way including the Bulge led to global reddening and stellar crowding maps \citep[][]{gonzalez2012, gonzalez2013, nidever2012, schultheis2014, Nataf2016, Surot2020, Nogueras-Lara2021, zhang2022, Sanders2022}, from which more detailed and accurate morphology and structural parameters of the Bulge could be derived \citep{wegg2013, portail2017,  simion2017, Clarke2019}. It is now established that in addition to the thick bar \citep{Stanek1994, Babusiaux2005}, the Galactic Bulge also presents the X-shape \citep{McWilliam+Zoccali2010, Nataf2010, gonzalez2015, Ness2016} in a composite structure that is challenging to unravel due to many overlapping components \citep{Zoccali+Valenti2016, Kunder2016, Lucey2021, Wylie2022, Marchetti2022, Rix2022}.

The different stellar populations in the Bulge exhibit different kinematics. Proper motions are thus a valuable tool to distinguish Bulge and disc stars and identify structures like streams or globular clusters \citep{Horta2021,Kader2022,Garro2022}, and to characterise structures like the nuclear star cluster or the nuclear stellar disc \citep{Clarke2022,Shahzamanian2022,NoguerasLara2022}. 

The recent near-infrared (NIR) surveys have revolutionised our understanding of the Galactic Bulge. The Vista Variables in the Via Lactea (VVV) and its extension, the VVV eXtended survey (VVVX), created maps of the Galactic Bulge and southern part of the disc ($-130\deg<l<20\deg$ and $-15\deg<b<10\deg$) covering $\sim1700\deg^2$ in five NIR passbands: $Z(0.87~\mu m)$, $Y(1.02~\mu m)$, $J(1.25~\mu m)$, $H(1.64~\mu m)$ and $K_s(2.14~\mu m)$ \citep{Minniti2010,Saito2012,AlonsoGarcia2018}. The VVV observations started in 2010 and continued with VVVX until 2022 collecting many tens of epochs in the $K_S$ band, and in the central parts, some areas have $>100$ epochs, enabling proper motion measurements over more than 10 years baseline. In the following, we use VVV to denote observations within the central Bulge area that were collected within the original VVV survey as well its extension VVVX.

VIRAC2 (Smith et al., in prep.) is the second data release of the VVV Infrared Astrometric Catalogue (VIRAC) \citep{Smith2018}. VIRAC2 is 90\% complete up to $K_S\sim16$ across the VVV bulge area \citep{Sanders2022}. Its proper motions are anchored to \textit{Gaia} absolute reference frame.

In addition to the NIR surveys, \textit{Gaia} presents an impressive amount of data that is a valuable source for studies of all galactic components.
The third data release (DR3) of the \textit{Gaia} survey \citep{GaiaDR3} 
contains, amongst other columns, the astrometric and photometric data for $\sim1.5$ billion sources spanning 34 months of observations \citep{Gaia_BrownEDR3}. \textit{Gaia} DR3 astrometry is in the International Celestial Reference System (ICRS) with the 2016.0 reference epoch. 
However, the survey has a completeness bellow 60\% for sources at $G\sim19$ or fainter and in stellar densities of about $5\times10^5\,\mathrm{stars}\,\deg^{-2}$, which happens in crowded fields like those within globular clusters. The completeness of \textit{Gaia} DR3 is below 20\% even for brighter sources in fields with stellar densities of about $10^6-10^7\,\mathrm{stars}\,\deg^{-2}$, which includes also the Galactic Bulge \citep{GaiaEDR3_Fabricius2021,EDR3SelFunct2022MNRAS.509.6205E,DR3SelFunct2022arXiv220809335C}.

As for any type of measurement, the \textit{Gaia} data are subject to systematics
coming from the instrument itself or from the data processing. The systematics affect both photometry and astrometry. The systematic offset in parallax or proper motion (i.e., zero points) can be obtained by a comparison with sources whose parallax and proper motion are known, like quasars or binaries. The zero-point can depend on magnitude, colour and position \citep[see][]{Lindegren2021,Riello2021}. 
The systematics could lead to the over or underestimation of measurements and their uncertainties.

The underestimation of the uncertainty estimates for positions, parallaxes and proper motions (i.e., the astrometric solution) is a known caveat since \textit{Gaia} DR1, where a formula for an inflation factor for the parallax formal uncertainties is provided \citep{Lindegren2016}. 
The underestimate of astrometric uncertainties of sources in crowded regions has been presented by \citet{Arenou_DR2val2018}. For \textit{Gaia} DR2, the formula for the parallax uncertainty underestimation factor includes both the systematics and the formal (catalogue) uncertainties \citep{Lindegren2018}. In DR2 the underestimation factor for parallax is between 2-3 for most sources. While DR3 improves that, the parallax formal uncertainties are underestimated for faint stars and stars in crowded fields \citep[see Sec. 3.11 in ][]{GaiaEDR3_Fabricius2021}; these stars might present also a larger underestimation in their proper motion uncertainties.

Several studies have investigated the accuracy of \textit{Gaia} proper motion and parallax measurements leading to correction factors \citep[e.g.,][]{GaiaEDR3_Fabricius2021,ElBadry2021,2021MNRAS.505.5978V,MaizApellaniz2021,Babusiaux2022}. Here we provide an independent study that has been triggered by our search for Hypervelocity stars in the inner Bulge \citep{Luna2019}. Hypervelocity stars are objects unbound from the Galactic potential and are extremely rare. Aiming to identify outliers in a large distribution, it is particularly important to know how reliable the measurements and their errors are. In this study, we aim to validate the use of \textit{Gaia} DR3 and VVV astrometry in crowded fields by comparing them to different sets of accurate and precise proper motions of HST observations.

\section{The data sets and catalogues cross-match}

The measurements of the stellar positions based on the HST photometry are very precise. The ACS/WFC plate scale of 50 $\mathrm{mas~pixel^{-1}}$ can result in proper motion measurements as accurate as $0.3 \mathrm{mas\,yr^{-1}}$ with observations spanning 2 years baseline \citep{Clarkson_BTS2008}. The plate scale of WFC3 UVIS channel is 40 $\mathrm{mas~pixel^{-1}}$ enabling similar precision for proper motion measurements \citep{BTS_Brown2009}. With more extended observing baseline over 9 years \citet{Calamida_SWEEPS2014} reported proper motion measurements accuracy of the order of $0.1 \mathrm{mas\,yr^{-1}}$ for the stars as faint as $F606W \sim 25.5$~mag combining data from both cameras. 
For comparison, the typical precision\footnote{\href{https://www.cosmos.esa.int/web/gaia/science-performance}{https://www.cosmos.esa.int/web/gaia/science-performance}} of \textit{Gaia} DR3 proper motions is $0.1 \mathrm{mas\,yr^{-1}}$ at $G=18$ and $0.5 \mathrm{mas\,yr^{-1}}$ at $G=20$.
\textit{Gaia} DR3 proper motions are superior for brighter stars, however, those are close to the saturation of the WFC/ACS detector in the HST datasets that cover the dense Bulge fields with multi-epoch observations. A more detailed comparison between the two instruments is given in \citet{delPino_GaiaHUB2022,Massari2020}.

As a benchmark, we use the \textit{WFC3 Galactic Bulge Treasury Program: Populations, Formation History, and Planets} (BTP)\footnote{Version 2 high-level science products from GO-11664.} \citep[][GO-11664; PI: Brown, T. M.]{Clarkson_BTS2008,BTS_Brown2009}, that consists of observations of four low reddening windows that map different environments in the Galactic Bulge: The SWEEPS, Stanek, Baade and Ogle29 windows (see Table \ref{tab:BTP location} for their location and extinction). The BTP was created to study resolved stellar populations using the WFC3 on the HST. The BTP provides a deep astrometric and photometric catalogue, reaching F606W$\sim 26$, several magnitudes below the Main Sequence turn-off (MSTO), around F606W$\sim 20$. The stars included in the BTP are field stars, as there are no stellar clusters in those fields of view (FoV).

\begin{table*}[]
    \centering
    \begin{tabular}{lccccc}
    \hline\hline
         Field & l (deg) & b (deg) & $A_v$ (mag)  & $N_f$ \textit{Gaia}-HST  & $N_f$ VIRAC2-HST \\
         \hline
          SWEEPS & $ +1.26$ & $ -2.65$ & $ 2.0$    & $97$  & 557\\
          Stanek & $+0.25$ & $ -2.15$ & $ 2.6$   & $ 110$  & 1559 \\
          Baade & $+1.06 $ & $ -3.81$ & $ 1.6$  & $ 30$  & ---\\
          Ogle29 & $-6.75 $ & $-4.71 $ & $ 1.5$  & $227 $   & 171 \\
          $\omega\,$Cen F1 & $-51.12 $ & $ +14.81 $ & $ 0.4$   & $369 $  & ---\\
          $\omega\,$Cen F2 & $-51.04 $ & $ +14.84 $ & $ 0.4$  & $817 $  & ---\\
          $\omega\,$Cen F3 & $ -51.11 $ & $ +15.01 $ & $0.4 $  & $982 $  & ---\\
          NGC 6652 & $ +1.53 $ & $ -11.38 $ & $ 0.37 $  & $978 $ & ---\\
         \hline 
    \end{tabular}
    \caption{Galactic coordinates, extinction coefficient, and the final number of matched sources ($N_f$) used for the analysis of the four BTP low-reddening windows, the outskirt fields of $\omega\,$Cen: F1, F2 and F3; and the globular cluster NGC 6652.}
    \label{tab:BTP location}
\end{table*}

To cross-match the HST sources with \textit{Gaia} DR3 and then compare their proper motions, we first transform the BTP proper motions, provided in the catalogue in units of pixel per baseline, into proper motions components along RA and DEC using a 30 $\mathrm{mas~pixel^{-1}}$ scale, as reported in the BTP catalogue description\footnote{\href{https://archive.stsci.edu/prepds/wfc3bulge/}{https://archive.stsci.edu/prepds/wfc3bulge/}} \citep{BTS_Brown2009}. We note that the images and proper motions are oriented North-East. The baseline for the SWEEPS window is of 6.2 years, with observations from 2004 to 2010, while for Stanek, Baade and Ogle29 is of two years with observations from 2010 to 2012. The larger baseline for the SWEEPS window is thanks to a previously observed programme in addition to the BTP \citep{Sahu2006,Clarkson_BTS2008}.

Once we computed the proper motions in $mas\,yr^{-1}$, we cleaned the data set examining the quality flags in the catalogues. The BTP positions and proper motions come from a PSF fit that uses three different photometric methods, depending on the brightness of the source. Within the BTP catalogue, the parameter \texttt{qual} indicates the fit quality and ranges between 0 and 1; we select the sources with the best quality fit, hence the best-characterised photometry and astrometry, restricting $0.9\leq$\texttt{qual}$\leq 1$. Such selection excludes bright sources close to the saturation of the WCF/ACS detector.

For \textit{Gaia} DR3, following \citet{Battaglia2022}\footnote{The authors propose an additional cut: the absolute value of the corrected excess factor (\texttt{phot\_bp\_rp\_excess\_factor}) within $5\sigma$ at the corresponding $G$-band magnitude \citep[see Eqs. 6 and 18 of ][]{Riello2021}. However, this cut pertains to photometric reliability, and since we deal with astrometry in this study, we decided not to take it into account in order to increase the statistics.}, we selected sources with a complete astrometric solution (\texttt{astrometric\_params\_solved} $\geq$ 31), not flagged as a duplicated source, 
and having a re-normalized unit weight error (\texttt{RUWE}) $<1.4$ \citep{Lindegren2018}; where RUWE is an indicator of the quality of the astrometric fit and that threshold is set to select single sources with good astrometric measurements. It has to be noted that in crowded fields, RUWE values may be underestimated, so one could in principle relax the requirement of RUWE$<1.4$ and thus increase the statistics.

The positions of stars in the BTP catalogue, given in equatorial coordinates (J2000 epoch), are propagated to 2016.0, the \textit{Gaia} DR3 reference epoch. 
We do an initial cross-match within $0\farcs5$ tolerance in position using the \texttt{astropy}  package \texttt{SkyCoord.match\_to\_catalog\_sky}, which uses a KD-Tree to find the nearest-neighbour. The HST astrometric reference frame has an offset $\leq 0\farcs3$ with respect to \textit{Gaia} DR1 \citep{Clarkson2018} in each coordinate, and a similar offset may still be present in \textit{Gaia} DR3 \citep{KozhurinaP2021}. To take into account the presence of systematic effects, we searched for trends between the proper motions and other parameters such as \textit{Gaia} DR3 colours, magnitude or position \citep[e.g.,][]{Massari2020,Massari2018}; however, there is no clear trend with any of the parameters.

\begin{figure}
\centering
\includegraphics[width=0.4\textwidth]{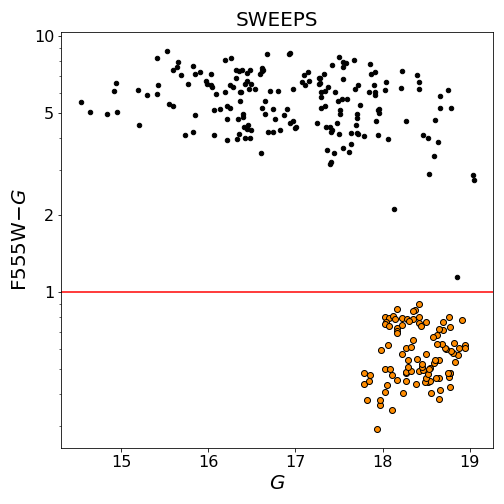}
\caption{\textit{Gaia} DR3 $G$-band and HST F555W magnitude difference. The initial position cross-match is refined with a selection of stars that have similar magnitudes. The good matches are below the red solid line at $\mathrm{F555W}-G=1$.} 
\label{fig:xmtch_mag_sweeps}
\end{figure}

Then, the cross-match is refined by comparing the source $G$-band magnitude and the HST F555W (or F606W) magnitude, leaving out the spurious matches. Figure \ref{fig:xmtch_mag_sweeps} shows the magnitude difference for the SWEEPS window. The correct matches lie below the red line. The cross-match is done individually in the different studied fields and the separation distribution of cross-matched sources in each field is centred at $\sim 0\farcs2$ with a standard deviation of $0\farcs4$.

Throughout the text, the comparison between \textit{Gaia} DR3 and the BTP windows is illustrated using the BTP-SWEEPS window, with its corresponding figures. The same methodology is applied for all BTP windows and their figures can be found in Appendix \ref{sec:appendixBTP}. In the case of Baade's window, there are 371 crossmatched sources within $0\farcs5$, but most of them are spurious matches, with a large $G$-band and F555W magnitude difference. Given that the final sample of well-matched sources in Baade's window consists of only 30 sources we will not consider this field in the rest of this work. 

\begin{figure}
    \centering
    \includegraphics[width=0.45\textwidth]{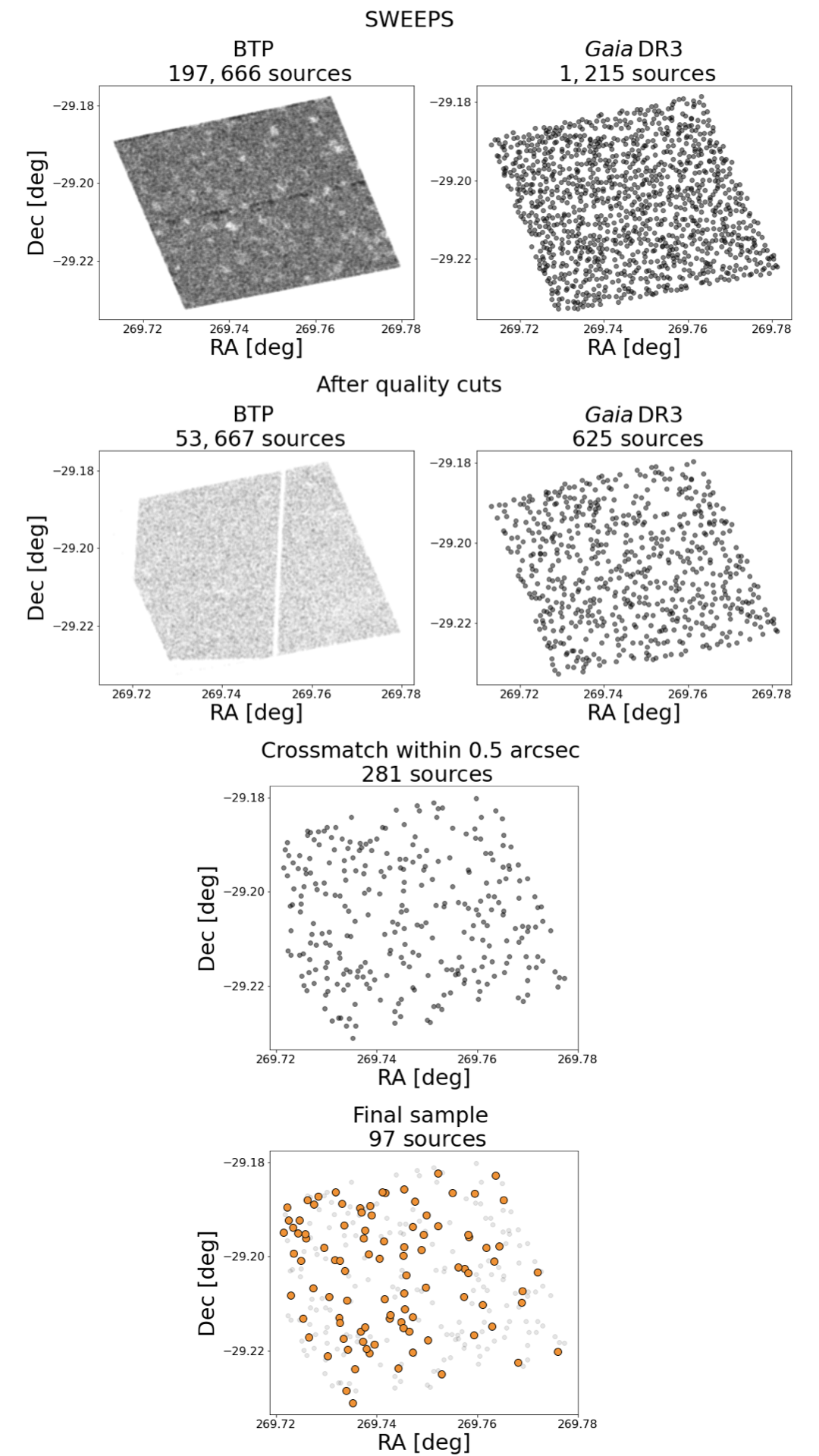}
    \caption{Number of sources in \textit{Gaia} DR3 and HST-BTP in the SWEEPS field. From top to bottom: initial sample in the same FoV, after applying the quality cuts described in Sect. 2, after a cross-match between the data sets with a tolerance of $0\farcs5$, and the final sample used for the study. The position of the sources in the final sample is plotted in orange. In the second row, the lack of sources in the corners of the BTP data set is because they did not have a proper motion value in the BTP catalogue. This is due to small differences in the field orientation between different epoch exposures.} 
    \label{fig:flow chart}
\end{figure}

Although the main HST data sets have $10^4-10^5$ sources, with $10^3$ \textit{Gaia} DR3 sources in the same FoV, after the cross-match, 
the final data set has only a few $10^2$ sources per field. Figure \ref{fig:flow chart} illustrates how the number of sources decreases as we apply the different quality cuts and cross-match refinements. From top to bottom of Fig. \ref{fig:flow chart}, sources from \textit{Gaia} DR3 catalogue are plotted on the right panels, while sources in the HST-BTP-SWEEPS window are plotted on the left panel. The panels in the second row
show the sources after the quality cuts described in the previous paragraphs. The final two panels show, in grey, the cross-matched sources within $0\farcs5$ of tolerance in position, and in orange in the bottom panel, the final sample used for the study. The orange points correspond to the orange points in Fig. \ref{fig:xmtch_mag_sweeps}. The number of sources remaining in each step of this cross-matching procedure for the HST-BTP-SWEEPS field is shown above each panel in Fig. \ref{fig:flow chart}.

\begin{figure}
\centering
\includegraphics[width=0.5\textwidth]{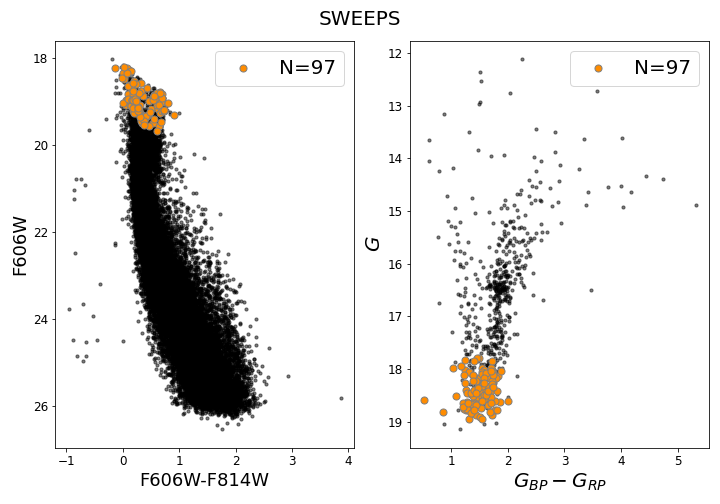}
\caption{Colour Magnitude Diagrams of \textit{Gaia} DR3 and HST in the SWEEPS window. The orange points correspond to the location of the cross-matched sources after the quality cuts and a 3$\sigma$ clipping in the uncertainty normalised proper motion difference. The black dots are the total sources in the field after the quality cuts. Both catalogues overlap in a short magnitude range.}
\label{fig:cmds_sweeps}
\end{figure}

The low number of final cross-matched sources is largely a consequence of the surveys differing in depth; while HST reaches magnitudes well below F606W$=24$, detecting mostly the main sequence stars, \textit{Gaia} is limited to stars brighter than $G=20$ mag. In the bright end, \textit{Gaia} is more complete, however, HST begins to saturate, with its bright detection limit
around F606W=18 mag. This is illustrated in Fig. \ref{fig:cmds_sweeps} showing the Colour Magnitude diagrams (CMDs) of \textit{Gaia} DR3 and HST in the SWEEPS field.

In addition to the comparison of the HST and \textit{Gaia} proper motions in high-density fields in the Bulge using the BTP catalogue, we also compare the HST and \textit{Gaia} DR3 proper motions in three fields in the outskirts of $\omega\,$Cen \citep[GO-14118 and GO-14662; PI: Bedin, L. R. Catalogues:][]{Bellini_WCENF12018,Scalco_WCENF2F32021,Libralato_F12018}, and in NGC\,6652 \citep[GO-13297; PI: Piotto, G. Catalogue:][]{Libralato_GCHST2022}, a globular cluster in Baade's window.

Finally, we also validate the proper motion uncertainties of the VIRAC2 catalogue with respect to the HST proper motions in three Bulge fields: SWEEPS, Stanek, and Ogle29 windows. The comparison is done for the $J$ band in VIRAC2 and F110W in HST. The data sets were cross-matched and cleaned adopting the same quality flags for \textit{Gaia} as described above. As a quality cut in VIRAC2, we select sources with a complete (5-parameter) astrometric solution, non-duplicates and that are detected in at least 20\% of the epochs. As an additional quality parameter, we select sources with unit weight error \texttt{uwe}$<1.2$, which is a threshold to select single sources with good astrometric measurements (L.~C.~Smith, private communication). The rest of the cross-match was done following a similar procedure as the one described for the \textit{Gaia}-BTP comparison. The cross-matched sources have a mean separation of $0\farcs25$. 
From the VIRAC2 initial sample of 5199, 5623, and 3856 in SWEEPS, Stanek, and Ogle29 windows, respectively, the final sample used for the comparison resulted in 557 in the SWEEPS field, 1515 sources in the Stanek field, and 171 in the Ogle29 field. The larger number statistics in SWEEPS and Stanek fields are thanks to the better matching magnitude range between the VVV and \textit{Gaia} DR3. The $H-J$ vs $J$ CMDs of the initial and final data sets are in Appendix \ref{sec:appendix_cdm_VIRAC2}.

\section{Inflation factor}

Assuming the proper motions in two different, uncorrelated, catalogues $\mu_1$ and $\mu_2$ with their corresponding uncertainties $\sigma_1$ and $\sigma_2$, and follow a Gaussian distribution, their uncertainty normalised difference ($\Delta \mu/\sigma_{\Delta \mu}$) should follow a normal distribution.

If the proper motion and/or its uncertainties in one of the catalogues are under or overestimated, the true proper motion uncertainty would be 

\begin{equation}
    \sigma_e^2=r^2\sigma_1^2
\end{equation}

\noindent with $r$ being the inflation factor needed to bring $\mu_1$ into a $1\sigma$ agreement with $\mu_2$.

Then, the variance of the uncertainty normalised proper motion difference is 

\begin{equation}
    K^2=\mathrm{Var}\left( \frac{\mu_1-\mu_2}{\sqrt{\sigma_1^2+\sigma_2^2}}\right)=\frac{\sigma_e^2+\sigma_2^2}{\sigma_1^2+\sigma_2^2}
\label{eq:K}
\end{equation}

Thus

\begin{equation}
    r^2=\frac{K^2(\sigma_1^2+\sigma_2^2)-\sigma_2^2}{\sigma_1^2}
\label{eq:inf_factor}
\end{equation}

Where $r$ is the individual inflation factor for each star in the sample which depends on the individual uncertainties $\sigma_1$ and $\sigma_2$, and the standard deviation $K$ of the uncertainty normalised proper motion difference of the parent population. 

In our case, we take $\mu_1$ and $\sigma_1$ as the reported \textit{Gaia} measurements, and $\mu_2$ and $\sigma_2$ as the HST measurements, where we assume that the HST proper motions measurements are true values with well-determined uncertainties that are neither under- nor over-estimated. Since we know this is not the case for the BTP catalogue, we looked for systematics in the HST data as discussed for example in \citet{Bellini2011} and \citet{Libralato_GCHST2022}, but find none. 
For globular clusters, we use HST proper motions that have been corrected for HST systematics (see Sec. \ref{sec:wcen} and \ref{sec:n6652}).

In this way, we incorporate the proper motion uncertainties underestimation completely into \textit{Gaia} DR3, even though HST proper motions are not free of systematics. Our goal is to set an upper limit on \textit{Gaia} DR3 proper motion uncertainties true underestimation.
To complement this, we include in Appendix \ref{sec:appendixBTP_rvar} an analysis of how the inflation factor changes when additional systematics to HST proper motions are introduced, and we show that for the SWEEPS window, $r$  varies in 20\% within the sampled range, however for Stanek and Ogle29 windows, $r$ decreases up to 50\% with larger PM uncertainty.
The derivation of the inflation factor was made independently for the proper motion components in RA and Dec. Therefore, the correlation parameter between RA and Dec components in \textit{Gaia} DR3 does not influence the analysis.

Any systematic errors in \textit{Gaia} values must be subtracted from the reported formal uncertainties to obtain the true uncertainties. Accounting for the systematic errors $\sigma_s$, the true \textit{Gaia} uncertainties would be 

\begin{equation}
    \sigma_e^2=r^2\sigma_1^2+\sigma_s^2
\end{equation}

\citet{2021A&A...649A...2L} derive \textit{Gaia} DR3 parallax and proper motion systematics through the angular covariance functions $V(\theta)$, which describe the spatial correlation of errors in astrometric quantities and depends on the angular separation, in degree, between two sources. They estimate the systematics from a quasar sample and find $V(\theta)_{\mu}\sim550~\mathrm{\mu as^2\,yr^{-2}}$, in the short-separation limit ($0\degree<\theta<0.125\degree$), which corresponds to a systematic error of $\sigma_S=0.017~\mathrm{mas\,yr^{-1}}$. \citet{2021MNRAS.505.5978V} extend this analysis for shorter separations with stars in Galactic globular clusters and they find $V(\theta)_{\mu}\sim700~\mathrm{\mu as^2\,yr^{-2}}$ in the short-separation limit, which corresponds to a systematic error of $\sigma_S=0.026~\mathrm{mas\,yr^{-1}}$. These numbers indicate the maximum precision of the proper motion measurements. This is an improvement with respect to
DR2 where the systematic has been constrained to $0.066 ~\mathrm{mas\,yr^{-1}}$ for sources with $G>16$ in \textit{Gaia} DR2 \citep{Lindegren2018}. 
The value of $\sigma_s$ can be otherwise obtained as the zero-point of the proper motion of quasars or other absolute-proper motion reference frame tracers, which we don't have in this comparison. Hence, our approach can only constrain the inflation factor $r$, which absorbs part of the systematic errors.

Taking $\sigma_s$ into account, the individual inflation factor is given by

\begin{equation}
    r^2=\frac{K^2(\sigma_1^2+\sigma_2^2)-\sigma_2^2-\sigma_s^2}{\sigma_1^2}
\label{eq:inf_factor_syst}
\end{equation}

\noindent where $\sigma_s=0.026~\mathrm{mas\,yr^{-1}}$ as the data sets areas are within the short-separation limit.

\section{Results of the fit}

In this section, we present the procedure and results of a Gaussian mixture model (GMM) fit to the uncertainty normalised proper motion difference. The same GMM fitting procedure that is applied to the BTP data set, is then also applied to assess the consistency of proper motion for other data sets with respect to \textit{Gaia} DR3. 

\subsection{BTP}

Due to the lack of sources in the Bulge that can be used to transform the proper motions to an absolute reference frame, the proper motions of the BTP fields are relative to the median of a sample of Bulge stars in the MSTO and fainter \citep{Clarkson2018}.
We transform the proper motions into absolute proper motions anchored to \textit{Gaia} by subtracting the mean of the $\mu_{\textit{BTP}}-\mu_{Gaia \mathrm{~DR3}}$ proper motion distributions. 
The BTP catalogue does not provide individual proper motion uncertainties, but an upper limit of $0.3~ \mathrm{mas~yr^{-1}}$, which we assume as the uncertainty value for the cross-matched sources. To assess the impact of this assumption on the results of our study,
we explore the extended SWEEPS catalogue \citep{Calamida_SWEEPS2014}, which contains individual proper motion uncertainties for all measured sources.
The results from the analysis of the extended SWEEPS catalogue are consistent with those of the SWEEPS-BTP window.
This is further discussed in Sect.\ref{sec:sweeps_full} with the corresponding plots in Appendix \ref{sec:appendixSWEEPSext}.

\begin{figure}
\centering
\includegraphics[width=0.5\textwidth]{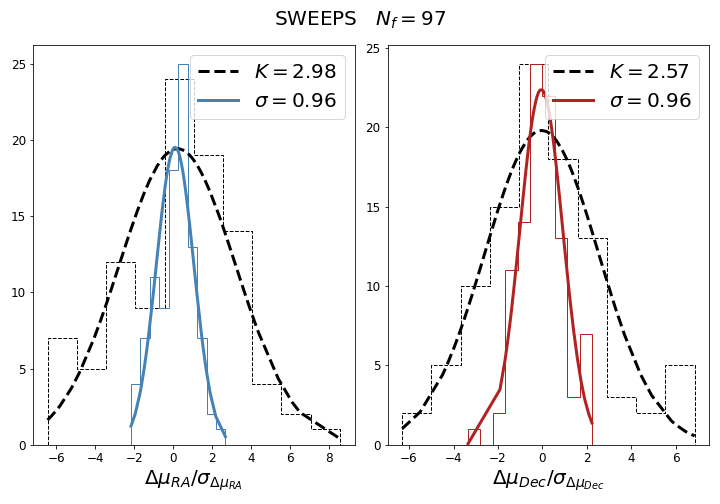}
\caption{Normalised proper motion difference distribution for the SWEEPS field. The dashed curves are the initial distributions in each coordinate component. The solid curves are the distributions after \textit{Gaia} DR3 proper motion uncertainties are multiplied by the inflation factor $r$. The standard deviation of the distribution before and after applying the inflation factor is labelled as $K$ and $\sigma$ respectively.}
\label{fig:pm_comparison}
\end{figure}

We compare the uncertainty normalised distributions of proper motion differences separately in RA and Dec and fit them with a Gaussian (see Fig. \ref{fig:pm_comparison}), keeping only the stars within $3\sigma$ of the distribution in each coordinate.
With $K$ being the standard deviation of $\Delta \mu /\sigma_\mu$, we apply Eq. \ref{eq:inf_factor} independently in each coordinate, to obtain the individual inflation factor $r$ of each source to be multiplied by its \textit{Gaia} proper motion uncertainty.

In Fig. \ref{fig:pm_comparison} we show $\Delta \mu /\sigma_\mu$ in RA (left panel) and Dec (right panel) for the 97 sources in the final sample in the BTP-SWEEPS window. The standard deviation of the $\Delta \mu /\sigma_\mu$ distribution before and after applying the inflation factor are labelled as $K$ and $\sigma$, respectively. The value was 2.98 in RA and 2.57 in Dec and, after applying the median inflation factor, it goes to $\sim1$ in both cases. The dashed Gaussian corresponds to the initial $\Delta \mu /\sigma_\mu$ distribution, while the solid-coloured Gaussian corresponds to the distribution after the inflation of the \textit{Gaia} DR3 proper motion uncertainty. The equivalent of Fig. \ref{fig:pm_comparison} for the studied fields is in Appendix \ref{sec: pm comp}.

The standard deviation of the distribution after the inflation of the \textit{Gaia} DR3 uncertainties is not exactly one because we use the median value of $r$ instead of an individual value for each star. 

\begin{figure*}[]
\centering
\includegraphics[width=0.9\textwidth]{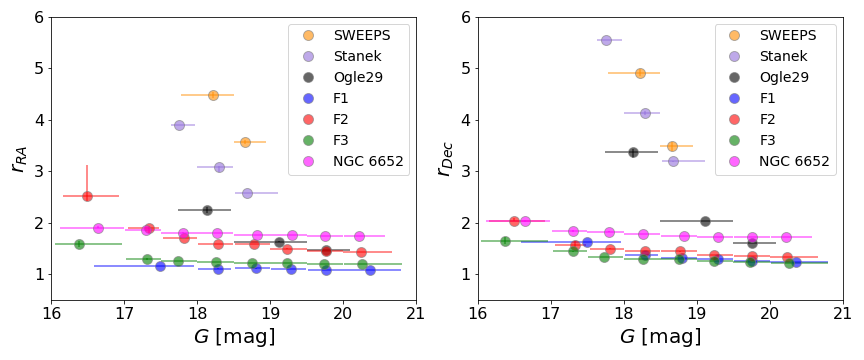}
\caption{Inflation factor $r$ dependence on $G$-band magnitude of the studied HST fields. The points represent the medians of the magnitude bins, and for $r$, the error bars are of the 16$^{th}$  and 84$^{th}$ percentiles. The bars in $G$ indicate the magnitude distribution at a given bin, where the marker is the median. The plotted data is in Table \ref{tab:r_G}.}
\label{fig:r_G_all_btsp}
\end{figure*}

The inflation factor $r$ depends strongly on the magnitude as shown in Fig. \ref{fig:r_G_all_btsp}, which is further described in Sect. 5.

\subsection{Extended SWEEPS catalogue}
\label{sec:sweeps_full}

\citet{Calamida_SWEEPS2014} published the extended baseline of observations of the SWEEPS window covering 9 years (GO-9750 and GO-12586; PI Sahu, K.C.), and derived proper motions in Galactic coordinates with precision around $0.1 \mathrm{mas\,yr^{-1}}$ at F606W=25.5 mag and around  $0.5 \mathrm{mas\,yr^{-1}}$ at F606W=28 mag. Based on such proper motions, they revealed the White Dwarf cooling sequence in the Galactic Bulge. We analysed this data set as well, because it offers independent validation of the results based on the BTP data, extending observations of the same field (SWEEPS) over a longer baseline and presenting individual uncertainties for all sources in the HST data set, rather than an average upper limit.

To compute the inflation factor and compare it with the results for the BTP windows, we have transformed the proper motions and their uncertainties from galactic coordinates into equatorial coordinates assuming that the proper motions are not correlated, as the catalogue does not come with a correlation coefficient.

The median inflation factor for the full-SWEEPS is consistent with the median inflation factor of the BTP-SWEEPS for every magnitude bin except for $G<18.5$ (see Appendix~\ref{sec:appendixSWEEPSext}) -- this bright end of the magnitude overlap between \textit{Gaia} DR3 and HST has few sources. Given the lack of quality flags in the full-SWEEPS catalogue, it is possible that few non-reliable astrometric measurements affect the sources in the bright bin.

\subsection{$\omega\,$Cen}
\label{sec:wcen}

The outer fields of $\omega\,$Cen (F1, F2 and F3) have a lower density than the BTP windows, in this case, the value of $r$ is closer to one (Fig. \ref{fig:r_G_all_btsp}), indicating that the uncertainties of the \textit{Gaia} DR3 proper motions are mostly consistent with those from the HST catalogues.
This shows that our procedure is consistent with previous findings such as e.g., \citet{MaizApellaniz2021,2021MNRAS.505.5978V,Babusiaux2022}.

We tried to analyse the core of $\omega\,$Cen with the HST data provided in \citet{Bellini_WcenCore2017}. However, the severe crowding in its core prevents any well-behaved measurement with \textit{Gaia}.
The majority of \textit{Gaia} DR3 sources in that region have a RUWE$>1.4$, and are thus rejected by
the quality cuts.

\subsection{NGC\,6652}
\label{sec:n6652}

Recently, \citet{Libralato_GCHST2022} published a homogeneous photometric and astrometric catalogue of 56 globular clusters in the Galaxy.

NGC\,6652 is a globular cluster of particular interest for this study as it is located within Baade's window in the Bulge, its data have been processed following the same procedure as the outer $\omega\,$Cen fields. The inflation factor is larger than the one derived for the $\omega\,$Cen outskirt fields, but lower than the Bulge fields, indicating that their proper motions are better characterised than those of the BTP fields. (Fig. \ref{fig:r_G_all_btsp}).

HST globular cluster comparisons were also used to validate \textit{Gaia} DR2 astrometry \citep{Arenou_DR2val2018} and DR3 \citep{GaiaEDR3_Fabricius2021}. The normalised dispersion of the differences in $\omega\,$Cen is close to 1, indicating that the uncertainties are correctly estimated \citep{Arenou_DR2val2018}.

\subsection{VIRAC2}

\begin{figure}
\centering
\includegraphics[width=0.5\textwidth]{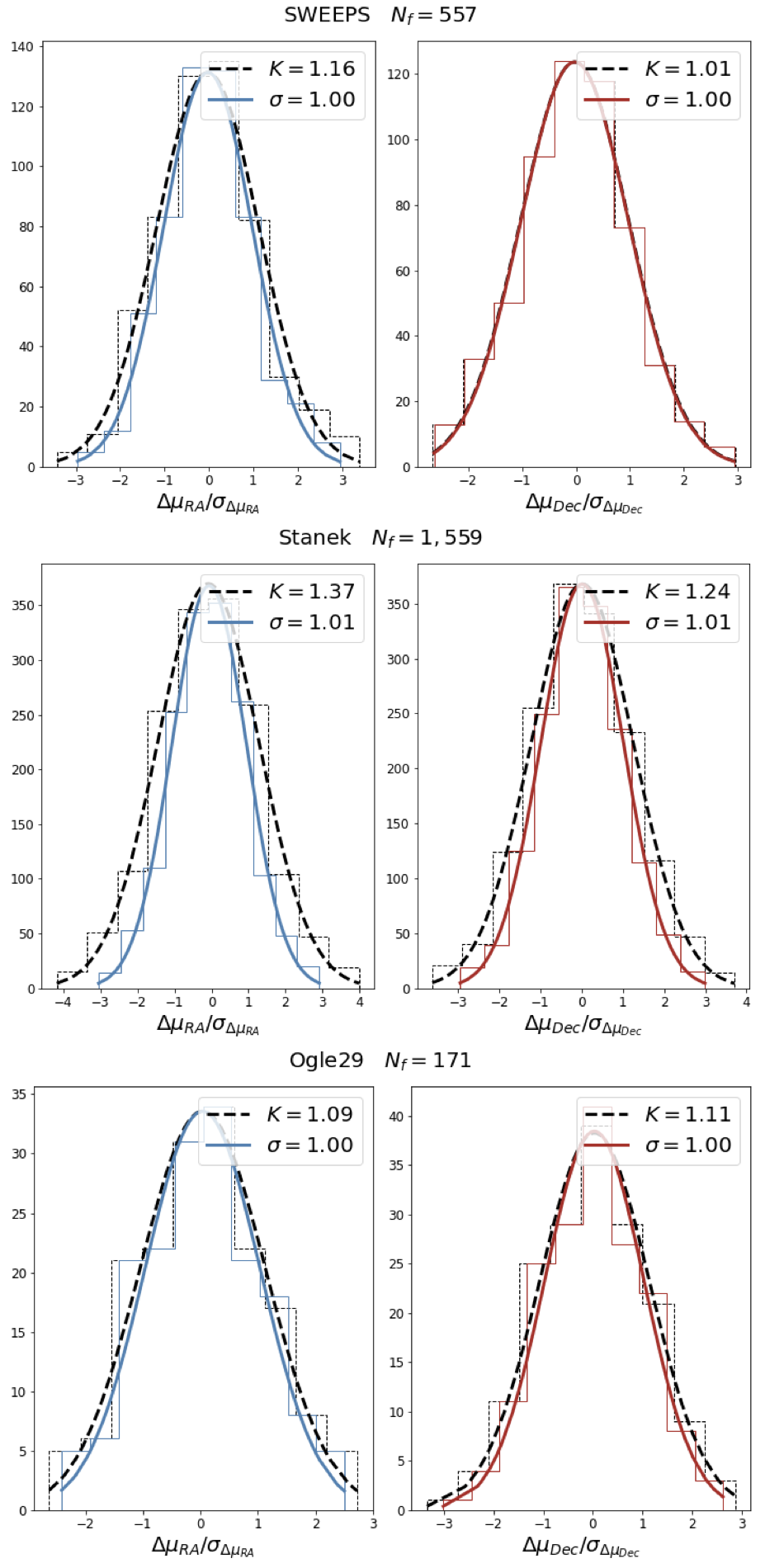}
\caption{Normalised proper motion difference distribution for VIRAC2 vs.\, SWEEPS-BTP (top), VIRAC2 vs.\ Stanek-BTP (middle), and VIRAC2 vs.\ Ogle29-BTP (bottom). The dashed curves are the initial distributions in each coordinate component. The solid curves are the distributions after VIRAC2 uncertainties are multiplied by the inflation factor $r$.}
\label{fig:pm_comparison_virac}
\end{figure}

As VIRAC2 covers the inner part of the Bulge, we have data for three of the seven studied fields: SWEEPS, Stanek and Ogle29 windows. VIRAC2 proper motions are in equatorial coordinates and anchored to \textit{Gaia} DR3 absolute reference frame. Hence, we transform the HST-BTP proper motions into absolute proper motions in the same manner as before, subtracting the mean of the $\mu_{BTP}-\mu_{VIRAC2}$ proper motion distributions independently in RA and Dec. For the BTP uncertainties, we assume an upper limit of 0.3 mas $yr^{-1}$.
Figure \ref{fig:pm_comparison_virac} is equivalent to Fig. \ref{fig:pm_comparison}, where the original $\Delta\mu/\sigma_{\mu}$ in each coordinate are plotted as dashed Gaussians and the solid Gaussians are the distributions after the inflation of VIRAC2 proper motion uncertainties. The top panels correspond to the comparison of VIRCAC2 vs HST in the BTP-SWEEPS field, the middle panels to the VIRAC2 and the BTP-Stanek comparison, and the bottom panels correspond to the comparison with the BTP-Ogle29 field. 

\section{Analysis and discussion}

In the catalogue validation of \textit{Gaia} DR3, \citet{GaiaEDR3_Fabricius2021}, find an offset in the unit weight error of parallax with respect to their uncertainty. The offset for the bulk of the sources is 1.05 for a 5-parameter (positions, parallax, and proper motions) astrometric solution and 1.22 for a 6-parameter astrometric solution, where the sixth parameter is the "pseudocolour", the astrometrically estimated effective wave number when the source colour $G_{BP}-G_{RP}$ was unavailable. Sources with a 6-parameter solution are more prone to systematics and spurious solutions \citep{Lindegren2021, GaiaEDR3_Fabricius2021}. There is an extra underestimation factor that depends on 
$G$-band magnitude \citep[Fig. 20 and 21 in][]{GaiaEDR3_Fabricius2021}.

\citet{delPino_GaiaHUB2022} applied the offsets (1.05 or 1.22) derived specifically for parallax uncertainties as underestimation factors also to correct uncertainties for all astrometric parameters
(positions, parallax and proper motions) in \texttt{GaiaHUB}, a code that uses HST observations as an extra epoch for \textit{Gaia}. They showed that this resulted in improved proper motion accuracy for the case studies.

This is not the only case in which an inflation factor to correct for the underestimation of \textit{Gaia} uncertainties has been derived for different cases with successful results.
\citet{ElBadry2021} found that \textit{Gaia} DR3 parallax uncertainties are underestimated up to 50\% by comparing the parallax of binary components' in a catalogue they constructed assuming the pairs are bound. 
\citet{MaizApellaniz2021} derive inflation factors for parallax in globular clusters, by forcing the parallax distribution to be normal; the inflation factor is larger for stars in the magnitude range of $12<G<18$, finding similar values as \citet{GaiaEDR3_Fabricius2021}. Following the same procedure, \citet{Babusiaux2022} derive an inflation factor for the radial velocity uncertainty in a sample of open clusters. They fit a two-order polynomial to the factor as a function of magnitude ($G_{RVS}$) and effective temperature, providing in this way a formula for the uncertainties inflation.

Furthermore, the astrometric solution of a source may be unreliable depending on its position. \citet{Rybizki2022MNRAS} made a classifier for accurate astrometric solutions in \textit{Gaia} DR3, finding that the majority of sources in the direction of the Bulge have a spurious astrometric solution. They compare \textit{Gaia} DR3 proper motions with OGLE IV proper motions and find that the uncertainties are underestimated for either one or both of them. The scanning law could amplify the disturbance of a close neighbour, which would have a larger effect in a crowded field such as the Bulge, resulting in spurious astrometric solutions.

As a result of the selection of the best quality sources available and the difference in crowding of the studied fields, the final sample of our data sets covers different magnitude ranges, 
as can be seen in Fig. \ref{fig:r_G_all_btsp}.
For the different studied fields, Table \ref{tab:r_G} lists the median inflation factor and the number of sources (N) use to compute it in a given bin. The number of sources in each magnitude bin ranges between $\sim20$ in the brightest magnitude bins and $>100$ at $G<18$ for some fields such as the $\omega\,$Cen fields. To verify whether the low number statistics and the individual sources with outlying $r$ may affect the median value of $r$, we performed a bootstrap sampling.

\begin{table}
    \centering
    \begin{tabular}{lccc}
    \hline\hline
& & & \\
SWEEPS & & & \\
$G$-band magnitude & $r_{RA}$ & $r_{Dec}$ & N \\
\hline
17.78 $\leq G <$ 18.49 & 4.45 & 4.91 & 56 \\
18.50 $\leq G <$ 18.94 & 3.57 & 3.48 & 41 \\
\hline
\hline

& & & \\
Stanek & & & \\
$G$-band magnitude & $r_{RA}$ & $r_{Dec}$ & N \\
\hline
17.64 $\leq G <$ 17.97 & 3.90 & 5.56 & 12 \\
18.00 $\leq G <$ 18.49 & 3.06 & 4.13 & 52 \\
18.52 $\leq G <$ 19.11 & 2.59 & 3.20 & 46 \\
\hline
\hline

& & & \\
Ogle29 & & & \\
$G$-band magnitude & $r_{RA}$ & $r_{Dec}$ & N \\
\hline
17.74 $\leq G <$ 18.46 & 2.26 & 3.32 & 23 \\
18.50 $\leq G <$ 19.49 & 1.63 & 2.04 & 111 \\
19.50 $\leq G <$ 20.09 & 1.48 & 1.60 & 93 \\
\hline
\hline

& & & \\
$\omega\,$Cen F1 & & & \\
$G$-band magnitude & $r_{RA}$ & $r_{Dec}$ & N \\
\hline
16.59 $\leq G <$ 17.96 & 1.16 & 1.63 & 15 \\
18.01 $\leq G <$ 18.47 & 1.11 & 1.38 & 30 \\
18.51 $\leq G <$ 19.00 & 1.12 & 1.33 & 42 \\
19.01 $\leq G <$ 19.49 & 1.10 & 1.30 & 64 \\
19.51 $\leq G <$ 19.99 & 1.09 & 1.26 & 83 \\
20.00 $\leq G <$ 20.79 & 1.08 & 1.24 & 133 \\
\hline
\hline

& & & \\
$\omega\,$Cen F2 & & & \\
$G$-band magnitude & $r_{RA}$ & $r_{Dec}$ & N \\
\hline
16.16 $\leq G <$ 16.93 & 2.83 & 2.04 & 18 \\
17.06 $\leq G <$ 17.48 & 1.87 & 1.57 & 26 \\
17.53 $\leq G <$ 18.00 & 1.71 & 1.50 & 58 \\
18.02 $\leq G <$ 18.49 & 1.59 & 1.45 & 90 \\
18.51 $\leq G <$ 18.99 & 1.60 & 1.45 & 145 \\
19.00 $\leq G <$ 19.50 & 1.50 & 1.38 & 175 \\
19.51 $\leq G <$ 19.99 & 1.45 & 1.35 & 177 \\
20.00 $\leq G <$ 20.66 & 1.44 & 1.34 & 123 \\
\hline
\hline

& & & \\
$\omega\,$Cen F3 & & & \\
$G$-band magnitude & $r_{RA}$ & $r_{Dec}$ & N \\
\hline
16.05 $\leq G <$ 16.97 & 1.58 & 1.68 & 21 \\
17.03 $\leq G <$ 17.50 & 1.30 & 1.45 & 29 \\
17.51 $\leq G <$ 17.99 & 1.27 & 1.34 & 82 \\
18.00 $\leq G <$ 18.49 & 1.24 & 1.30 & 144 \\
18.50 $\leq G <$ 19.00 & 1.23 & 1.30 & 174 \\
19.01 $\leq G <$ 19.49 & 1.21 & 1.26 & 202 \\
19.50 $\leq G <$ 20.00 & 1.20 & 1.24 & 172 \\
20.01 $\leq G <$ 20.80 & 1.20 & 1.23 & 152 \\
\hline
\hline

& & & \\
NGC 6652 & & & \\
$G$-band magnitude & $r_{RA}$ & $r_{Dec}$ & N \\
\hline
16.12 $\leq G <$ 16.99 & 1.91 & 2.04 & 29 \\
17.01 $\leq G <$ 17.49 & 1.86 & 1.84 & 37 \\
17.50 $\leq G <$ 18.00 & 1.81 & 1.83 & 47 \\
18.01 $\leq G <$ 18.49 & 1.81 & 1.78 & 73 \\
18.50 $\leq G <$ 19.00 & 1.76 & 1.74 & 195 \\
19.01 $\leq G <$ 19.49 & 1.76 & 1.73 & 239 \\
19.50 $\leq G <$ 20.00 & 1.75 & 1.72 & 217 \\
20.01 $\leq G <$ 20.58 & 1.75 & 1.72 & 138 \\

    \hline
    \end{tabular}
    \caption{Median inflation factor ($r$) in a given $G$-band magnitude range for the seven studied fields. N is the number of sources used to compute the median inflation factor in a given bin (see Fig. \ref{fig:r_G_all_btsp}).}
    \label{tab:r_G}
\end{table}

To account for the influence of the low number statistics in the results and to estimate the uncertainty of the inflation factor derivation, we did a bootstrap sampling with 1000 iterations of $N-5$ elements, where $N$ is the number of sources in the parent sample. The error bars for $r$ in Fig. \ref{fig:r_G_all_btsp} are smaller than the marker in most cases.

\begin{figure*}
\centering
\includegraphics[width=0.9\textwidth]{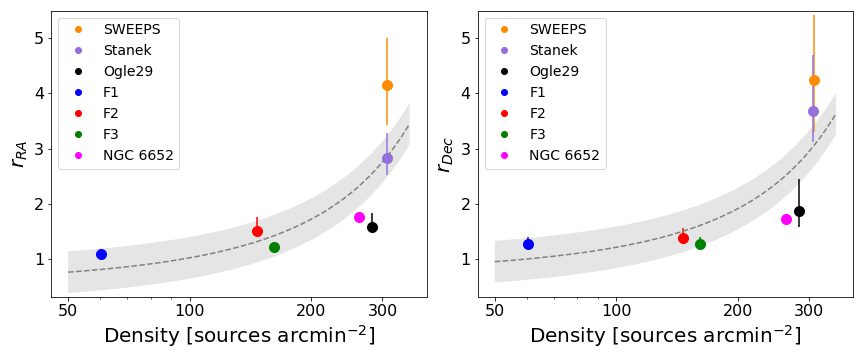}
\caption{Inflation factor $r$ dependence on the number density defined of sources found in the $\textit{Gaia}$ DR3 catalogue in a given field. The median $r$ for the different fields is plotted with different colours. The error bars correspond to the 16$^{th}$ and 84$^{th}$ percentile in the distribution of $r$ across the magnitude range of a given field. The shade is the 1$\sigma$ uncertainty on the fit. The plotted data are in Table \ref{tab:r_dens}.}
\label{fig:r_density_all_btsp}
\end{figure*}

Figure \ref{fig:r_density_all_btsp} shows the dependence of $r$ with respect to stellar surface density, where the proxy for the density is the number of \textit{Gaia} DR3 sources per arcmin$^2$ in a given field ($N$), that is, before any quality cut and for the entire magnitude range. \textit{Gaia} uncertainties are more underestimated as the stellar surface density increases, such as for SWEEPS and Stanek fields. This dependence follows the relation:

\begin{equation}
    r_{i}= 1.004\,N^{\alpha_i}
\label{eq: r_dens}
\end{equation}

\noindent where $i=$ RA, Dec; and $\alpha=-0.46 \pm 0.38$ in RA and $\alpha=-0.27 \pm 0.38$ in Dec.

The dependence of the inflation factor with density has a trend that follows Eq. \ref{eq: r_dens}; this trend is plotted as a grey dashed line in Fig. \ref{fig:r_density_all_btsp}. \citet{2021MNRAS.505.5978V} studied the parallax uncertainty inflation factor dependence on the stellar density of several Galactic globular clusters. They find that the parallax inflation factor follows an exponential function that depends on the stellar number density (see their Fig.~5). Although there is a 
similarity between the shape of their relation and ours, it is not directly comparable, because
our approach, in contrast to \citet{2021MNRAS.505.5978V}, compares \textit{Gaia} to other catalogues.
could expect that NGC\,6652, as a globular cluster in the Galactic Bulge, should have a higher stellar density than the Ogle29 field. However blending affecting \textit{Gaia} may result in a lower count of sources, hence the similar location of both fields in Fig. \ref{fig:r_density_all_btsp}.

\begin{table*}[]
    \centering
    \begin{tabular}{lcccc}
    \hline\hline
         Field & N(Gaia) & area (arcmin$^2$) & median $r_{RA}$ & median $r_{Dec}$ \\
    \hline
SWEEPS & 2773 & 9.01 & $4.15_{-0.73}^{+0.86}$ & $4.24_{-0.94}^{+1.18}$ \\ \\
Stanek & 2737 & 8.91 & $2.82_{-0.30}^{+0.46}$ & $3.68_{-0.56}^{+1.02}$ \\ \\
Ogle29 & 2774 & 9.82 & $1.57_{-0.10}^{+0.26}$ & $1.87_{-0.29}^{+0.58}$ \\ \\
$\omega\,$Cen F1 & 847 & 14.02 & $1.09_{-0.01}^{+0.03}$ & $1.26_{-0.03}^{+0.12}$ \\ \\
$\omega\,$Cen F2 & 2181 & 14.89 & $1.50_{-0.06}^{+0.25}$ & $1.38_{-0.03}^{+0.18}$ \\ \\
$\omega\,$Cen F3 & 2438 & 15.09 & $1.22_{-0.02}^{+0.05}$ & $1.27_{-0.04}^{+0.13}$ \\ \\
NGC 6652 & 3714 & 14.14 & $1.76_{-0.01}^{+0.04}$ & $1.73_{-0.01}^{+0.05}$ \\ \\
    \hline
    \end{tabular}
    \caption{Number of sources in \textit{Gaia} DR3 in a given field, area of the field, and median values of the inflation factor $r$, with the 16$^{th}$ and 84$^{th}$ percentile in $r$ indicated (see Fig. \ref{fig:r_density_all_btsp}).}
    \label{tab:r_dens}
\end{table*}

The difference between the inflation factor in RA and Dec can be explained as a result of the \textit{Gaia} scanning law \citep{Gaia_BrownEDR3} because the survey is more precise perpendicular to the ecliptic 
where the satellite has more transits. We note that the Ogle29 field presents the largest difference between the inflation factor values for RA and Dec when compared to other BTP windows. In this location, the Dec component is oriented close to being perpendicular to the ecliptic, thus being more precise. The median $\sigma_{Dec}$ is $0.25~\mathrm{mas\,yr^{-1}}$, while the median $\sigma_{RA}$ is $0.46~\mathrm{mas\,yr^{-1}}$.

The inflation factor $r$ shows a strong dependence on magnitude. As the precision in \textit{Gaia} DR3 increases for brighter sources, a larger $r$ is needed to bring  the uncertainty-weighted proper motions
into a 1$\sigma$ agreement with the HST measurements.
Figure \ref{fig:r_G_all_btsp} shows the behaviour of the median inflation factor vs. $G$-band magnitude: the median inflation factor increases towards the bright end. The higher density fields have overall higher inflation factor values, but the trend with magnitude is similar between fields.

Although in general, the inflation factor $r$ ranges between 1--2 in the lower density fields F1, F2 and F3 in $\omega\,$Cen, and in NGC\,6652 (see Fig. \ref{fig:r_G_all_btsp}), its value increases in the bright bin ($G<17$) in $\omega\,$Cen F2 field on RA. This is because of the low number of stars (18) in that magnitude bin, which is reflected by the larger error bar for $r$.

\begin{figure}
    \centering
    \includegraphics[width=0.4\textwidth]{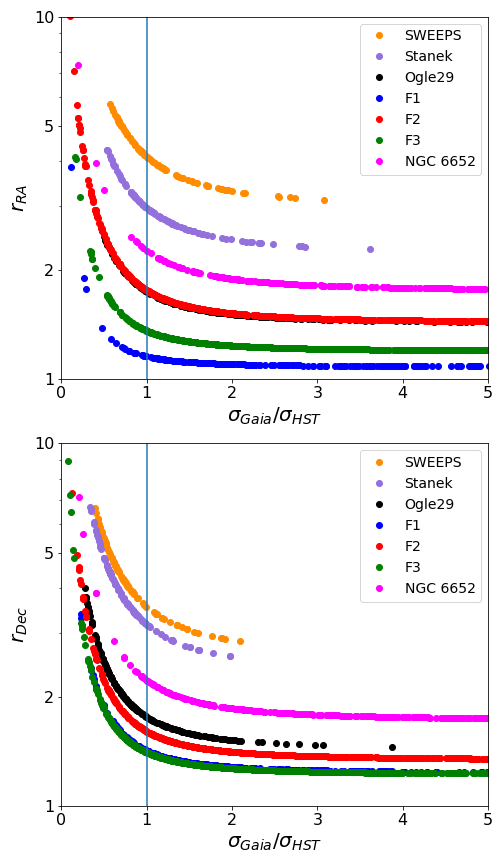}
    \caption{Inflation factor dependence on \textit{Gaia} and HST uncertainty ratio. The vertical line marks $\sigma_{Gaia}=\sigma_{HST}$. For sources with $\sigma_{Gaia}>\sigma_{HST}$, the inflation factor reaches asymptotically a minimum value which corresponds to the standard deviation $K$ of the parent population. This behaviour follows the functional form of Eq. \ref{eq:inf_factor}.}
    \label{fig:sigratio_r_all}
\end{figure}

As \textit{Gaia} DR3 measurements become more precise, and HST starts to reach the saturation point, the bright sources whose proper motions diverge from HST need a larger multiplicative factor to bring them into concordance. Figure \ref{fig:sigratio_r_all} shows how from the functional form of Eq. \ref{eq:inf_factor}, $r$ has a steep increase for sources with $\sigma_{Gaia}<\sigma_{HST}$.
The vertical line in Fig. \ref{fig:sigratio_r_all} represents the point when $\sigma_{Gaia}=\sigma_{HST}$. At the faint end ($G>17$), where $\sigma_{Gaia}>\sigma_{HST}$, the photon noise dominates the uncertainties over the systematics \citep{GaiaEDR3_Fabricius2021}. Therefore the inflation factor has an asymptotic behaviour tending to the standard deviation ($K$) of the parent population. 

\section*{Comparison of VIRAC2 and HST}

\begin{figure*}[]
\centering
\includegraphics[width=0.9\textwidth]{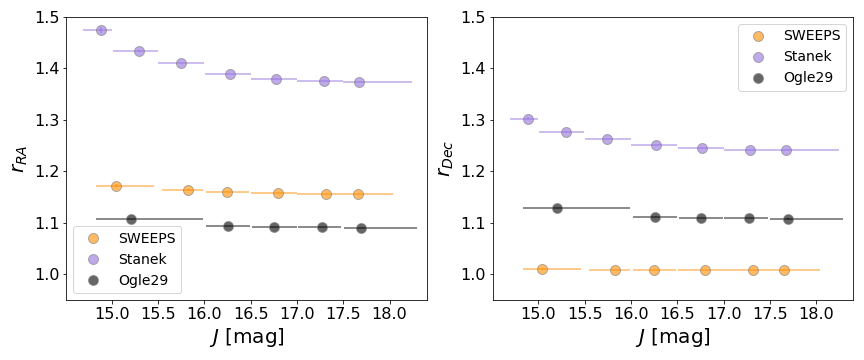}
\caption{Inflation factor $r$ dependence on VIRAC2 $J$ magnitude. The proper motion comparison was made between VIRAC2 and the HST-BTP SWEEPS, Stanek, and Ogle29 windows. The points are the median values of $J$ in each magnitude bin; the error bars indicate the magnitude distribution at a given bin. In $r$, the error bars are the 16$^{th}$ and 84$^{th}$ percentile.}
\label{fig:r_J_virac}
\end{figure*}

VIRAC2 proper motions have a precision comparable to those of the HST.
Table \ref{tab:r_J} lists the median inflation factor at a given $J$-band magnitude bin
shown in Fig. \ref{fig:r_J_virac}. 
Figure \ref{fig:r_J_virac} presents the dependence of the inflation factor $r$ as a function of $J$-band magnitude for the comparison of VIRAC2 with respect to the HST data in SWEEPS, Stanek and Ogle29 BTP fields. 
In contrast with the results for the inflation factor for the \textit{Gaia} DR3 proper motion uncertainties (Fig. \ref{fig:r_density_all_btsp}), the comparison with the Stanek window results in a larger inflation factor than the inflation factor for the SWEEPS window. Furthermore, within the range of magnitudes that can be probed, $r$ is nearly constant and does not depend on the brightness. In RA, VIRAC2 proper motion uncertainties in both fields are slightly underestimated, while the inflation factor in Dec is low, that is, close to one.
This analysis shows that the proper motions uncertainties of VIRAC2 and HST are consistent within $1\sigma$. 

\begin{table}
    \centering
    \begin{tabular}{lccc}
    \hline\hline
& & & \\
SWEEPS & & & \\
$J$-band magnitude & $r_{RA}$ & $r_{Dec}$ & N \\
\hline
14.83 $\leq J<$ 15.46 & 1.17 & 1.01 & 20 \\
15.55 $\leq J<$ 15.99 & 1.16 & 1.01 & 18 \\
16.02 $\leq J<$ 16.48 & 1.16 & 1.01 & 70 \\
16.50 $\leq J<$ 16.99 & 1.16 & 1.01 & 101 \\
17.00 $\leq J<$ 17.49 & 1.16 & 1.01 & 210 \\
17.50 $\leq J<$ 18.04 & 1.16 & 1.01 & 138 \\
\hline
\hline
& & & \\
Stanek & & & \\
$J$-band magnitude & $r_{RA}$ & $r_{Dec}$ & N \\
\hline
14.69 $\leq J<$ 15.0 & 1.47 & 1.3 & 31 \\
15.01 $\leq J<$ 15.5 & 1.44 & 1.28 & 91 \\
15.50 $\leq J<$ 15.99 & 1.41 & 1.26 & 163 \\
16.00 $\leq J<$ 16.5 & 1.39 & 1.25 & 231 \\
16.50 $\leq J<$ 16.99 & 1.38 & 1.24 & 300 \\
17.00 $\leq J<$ 17.49 & 1.38 & 1.24 & 419 \\
17.50 $\leq J<$ 18.24 & 1.37 & 1.24 & 324 \\
\hline
\hline
& & & \\
Ogle29 & & & \\
$J$-band magnitude & $r_{RA}$ & $r_{Dec}$ & N \\
\hline
14.83 $\leq J<$ 15.99 & 1.11 & 1.13 & 6 \\
16.02 $\leq J<$ 16.50 & 1.09 & 1.11 & 15 \\
16.52 $\leq J<$ 17.00 & 1.09 & 1.11 & 32 \\
17.01 $\leq J<$ 17.48 & 1.09 & 1.11 & 64 \\
17.51 $\leq J<$ 18.29 & 1.09 & 1.11 & 54 \\
    \hline
    \end{tabular}
    \caption{Median inflation factor ($r$) in a given $J$-band magnitude range for the seven studied fields. N is the number of sources used to compute the median inflation factor in a given bin (see Fig. \ref{fig:r_J_virac}).}
    \label{tab:r_J}
\end{table}

\section{Conclusions}

In this paper, we present an analysis of \textit{Gaia} DR3 and VIRAC2 proper motion uncertainties in comparison with the HST proper motions in seven 
fields that present different levels of crowding. Three fields are located in the direction towards the Galactic Bulge: the SWEEPS, Stanek and Ogle29 BTP windows; furthermore, we included three outskirt fields in the globular cluster $\omega\,$Cen, and the globular cluster NGC 6652. 
Our main findings concerning the comparison between \textit{Gaia} DR3 and HST data sets are summarised as follows:

\begin{itemize}
    \item  
    To bring the \textit{Gaia} DR3 proper motions 
    into a 1$\sigma$ agreement with those from the HST, we need an inflation factor to account for the underestimation of their uncertainties and the systematic errors. HST proper motions are not free of systematics, and part of the underestimation may well 
    belong to those. Since at the moment, we cannot quantify such underestimation in HST proper motions, we impose the underestimation to \textit{Gaia} as a way to extend its capabilities and take the inflation as an upper limit on the true underestimation.
    \item The inflation factor presents a dependence on stellar surface density, indicated by the number of sources in \textit{Gaia} DR3 catalogue per arcmin$^2$. For the dense BTP fields, there is also a strong dependence on $G$-band magnitude, which is driven entirely by the sources where \textit{Gaia} DR3 is more precise than HST. These are the brightest sources in the samples.
    \item The inflation factor dependence on stellar surface density follows an exponential function with a factor of $-0.46$ in RA and $-0.27$ in Dec (see Eq. \ref{eq: r_dens}).
    \item In the most crowded fields, such as the BTP windows, the inflation factor ranges from a factor of two for $G\sim19$, to a factor of five at $G<18$. 
    \item In less dense fields, such as the outskirts of the globular clusters $\omega\,$Cen and NGC\,6652, the inflation factor is less than two, indicating a better agreement of the measurements and their uncertainties.

\end{itemize}

The large inflation factor in the dense Bulge fields implies that the proper motion uncertainties for the brightest sources are underestimated either for the HST or for \textit{Gaia} DR3, while the proper motion measurements for the globular cluster fields present much better agreement between the HST and \textit{Gaia} DR3 catalogues. We note that all of the sources in the final sample of the BTP-\textit{Gaia} DR3 cross-match have a 6-parameter astrometric solution, whereas, for the $\omega\,$Cen outskirt fields F1, F2 and F3, the amount of 6-parameter sources is 38\%, 59\% and 60\%, respectively; for the globular cluster NGC\,6652, it is 62\%. This then supports the conclusion that the proper motion uncertainties underestimation is more likely affecting \textit{Gaia} than the HST measurements in the Bulge fields.

VIRAC2 is deeper and more complete than \textit{Gaia} DR3 in the BTP fields.
There is no need for an inflation factor as the VIRAC2 proper motion uncertainties are in good agreement with the HST ones. Therefore for these crowded Bulge fields, we recommend using VIRAC2 proper motions as a complement of \textit{Gaia} DR3 proper motions either to increase statistics or in case \textit{Gaia} DR3 proper motions have not a well-behaved astrometric solution.

The main limitation of this study was the low number of statistics
due to the different depths of the fields.
They typically had overlapping magnitude ranges where the quality of the measurements appears to be insufficiently characterised.
While we acknowledge the advantage of the NIR data for the highly extincted and crowded Galactic Bulge fields, \textit{Gaia} data are nevertheless very valuable. They provide an independent check on the proper motion measurements, potentially extend the baseline and magnitude range, as well as provide the possibility to study outliers in the proper motion distribution and thus search for hypervelocity stars, which was our original goal that motivated this study.
\textit{Gaia} proper motion uncertainties will improve, as the observational time baseline increases. A factor of $\sim2$ improvement can be expected in future releases\footnote{\href{https://www.cosmos.esa.int/web/gaia/science-performance}{https://www.cosmos.esa.int/web/gaia/science-performance}}. 
This will allow us to further characterise \textit{Gaia} vs. VIRAC2 astrometric uncertainties and exploit the full capabilities of both surveys.

\section*{Acknowledgements}

We are grateful to Anthony G.\,A. Brown for discussing and commenting on the early draft, and Leigh Smith for generously sharing the preliminary VIRAC2 data.  A.\,L. acknowledges support from the ANID Doctorado Nacional 2021 scholarship 21211520, and the ESO studentship. A.\,L. is grateful to T. Kolcu, Z. Penoyre, S. Verberne, F.\,A. Evans, E.\,M. Rossi, and J. de Bruyne for the useful discussions. D.\,M. gratefully acknowledges support by the ANID BASAL projects ACE210002 and FB210003, by Fondecyt Project No. 1220724, and by CNPq/Brazil through project350104/2022-0.
This work has made use of data from the European Space Agency (ESA) mission
{\it Gaia} (\url{https://www.cosmos.esa.int/gaia}), processed by the {\it Gaia}
Data Processing and Analysis Consortium (DPAC,
\url{https://www.cosmos.esa.int/web/gaia/dpac/consortium}). Funding for the DPAC
has been provided by national institutions, in particular the institutions
participating in the {\it Gaia} Multilateral Agreement.
The data used here may be obtained from \url{https://archive.stsci.edu/prepds/wfc3bulge/}.
This research made use of Astropy, a community-developed core Python package for Astronomy \citep{2018AJ....156..123A, 2013A&A...558A..33A}.
This research made use of NumPy \citep{harris2020array}, SciPy \citep{Virtanen_2020} and Scikit-learn \citep{scikit-learn}. Finally, we are grateful to the anonymous referee for the insightful evaluation of our work and useful suggestions.

\bibliographystyle{aa}
\bibliography{ref}

\begin{thebibliography}{87}
\expandafter\ifx\csname natexlab\endcsname\relax\def\natexlab#1{#1}\fi

\bibitem[{{Alonso-Garc{\'\i}a} {et~al.}(2018){Alonso-Garc{\'\i}a}, {Saito},
  {Hempel}, {Minniti}, {Pullen}, {Catelan}, {Ramos}, {Cross}, {Gonzalez},
  {Lucas}, {Palma}, {Valenti}, \& {Zoccali}}]{AlonsoGarcia2018}
{Alonso-Garc{\'\i}a}, J., {Saito}, R.~K., {Hempel}, M., {et~al.} 2018, \aap,
  619, A4

\bibitem[{{Arenou} {et~al.}(2018){Arenou}, {Luri}, {Babusiaux}, {Fabricius},
  {Helmi}, {Muraveva}, {Robin}, {Spoto}, {Vallenari}, {Antoja},
  {Cantat-Gaudin}, {Jordi}, {Leclerc}, {Reyl{\'e}}, {Romero-G{\'o}mez}, {Shih},
  {Soria}, {Barache}, {Bossini}, {Bragaglia}, {Breddels}, {Fabrizio},
  {Lambert}, {Marrese}, {Massari}, {Moitinho}, {Robichon}, {Ruiz-Dern},
  {Sordo}, {Veljanoski}, {Eyer}, {Jasniewicz}, {Pancino}, {Soubiran}, {Spagna},
  {Tanga}, {Turon}, \& {Zurbach}}]{Arenou_DR2val2018}
{Arenou}, F., {Luri}, X., {Babusiaux}, C., {et~al.} 2018, \aap, 616, A17

\bibitem[{{Astropy Collaboration} {et~al.}(2018){Astropy Collaboration},
  {Price-Whelan}, {Sip{\H o}cz}, {G{\"u}nther}, {Lim}, {Crawford}, {Conseil},
  {Shupe}, {Craig}, {Dencheva}, {Ginsburg}, {VanderPlas}, {Bradley},
  {P{\'e}rez-Su{\'a}rez}, {de Val-Borro}, {Aldcroft}, {Cruz}, {Robitaille},
  {Tollerud}, {Ardelean}, {Babej}, {Bach}, {Bachetti}, {Bakanov}, {Bamford},
  {Barentsen}, {Barmby}, {Baumbach}, {Berry}, {Biscani}, {Boquien}, {Bostroem},
  {Bouma}, {Brammer}, {Bray}, {Breytenbach}, {Buddelmeijer}, {Burke},
  {Calderone}, {Cano Rodr{\'{\i}}guez}, {Cara}, {Cardoso}, {Cheedella},
  {Copin}, {Corrales}, {Crichton}, {D'Avella}, {Deil}, {Depagne}, {Dietrich},
  {Donath}, {Droettboom}, {Earl}, {Erben}, {Fabbro}, {Ferreira}, {Finethy},
  {Fox}, {Garrison}, {Gibbons}, {Goldstein}, {Gommers}, {Greco}, {Greenfield},
  {Groener}, {Grollier}, {Hagen}, {Hirst}, {Homeier}, {Horton}, {Hosseinzadeh},
  {Hu}, {Hunkeler}, {Ivezi{\'c}}, {Jain}, {Jenness}, {Kanarek}, {Kendrew},
  {Kern}, {Kerzendorf}, {Khvalko}, {King}, {Kirkby}, {Kulkarni}, {Kumar},
  {Lee}, {Lenz}, {Littlefair}, {Ma}, {Macleod}, {Mastropietro}, {McCully},
  {Montagnac}, {Morris}, {Mueller}, {Mumford}, {Muna}, {Murphy}, {Nelson},
  {Nguyen}, {Ninan}, {N{\"o}the}, {Ogaz}, {Oh}, {Parejko}, {Parley}, {Pascual},
  {Patil}, {Patil}, {Plunkett}, {Prochaska}, {Rastogi}, {Reddy Janga},
  {Sabater}, {Sakurikar}, {Seifert}, {Sherbert}, {Sherwood-Taylor}, {Shih},
  {Sick}, {Silbiger}, {Singanamalla}, {Singer}, {Sladen}, {Sooley},
  {Sornarajah}, {Streicher}, {Teuben}, {Thomas}, {Tremblay}, {Turner},
  {Terr{\'o}n}, {van Kerkwijk}, {de la Vega}, {Watkins}, {Weaver}, {Whitmore},
  {Woillez}, {Zabalza}, \& {Astropy Contributors}}]{2018AJ....156..123A}
{Astropy Collaboration}, {Price-Whelan}, A.~M., {Sip{\H o}cz}, B.~M., {et~al.}
  2018, \aj, 156, 123

\bibitem[{{Astropy Collaboration} {et~al.}(2013){Astropy Collaboration},
  {Robitaille}, {Tollerud}, {Greenfield}, {Droettboom}, {Bray}, {Aldcroft},
  {Davis}, {Ginsburg}, {Price-Whelan}, {Kerzendorf}, {Conley}, {Crighton},
  {Barbary}, {Muna}, {Ferguson}, {Grollier}, {Parikh}, {Nair}, {Unther},
  {Deil}, {Woillez}, {Conseil}, {Kramer}, {Turner}, {Singer}, {Fox}, {Weaver},
  {Zabalza}, {Edwards}, {Azalee Bostroem}, {Burke}, {Casey}, {Crawford},
  {Dencheva}, {Ely}, {Jenness}, {Labrie}, {Lim}, {Pierfederici}, {Pontzen},
  {Ptak}, {Refsdal}, {Servillat}, \& {Streicher}}]{2013A&A...558A..33A}
{Astropy Collaboration}, {Robitaille}, T.~P., {Tollerud}, E.~J., {et~al.} 2013,
  \aap, 558, A33

\bibitem[{{Babusiaux} {et~al.}(2023){Babusiaux}, {Fabricius}, {Khanna},
  {Muraveva}, {Reyl{\'e}}, {Spoto}, {Vallenari}, {Luri}, {Arenou},
  {{\'A}lvarez}, {Anders}, {Antoja}, {Balbinot}, {Barache}, {Bauchet},
  {Bossini}, {Busonero}, {Cantat-Gaudin}, {Carrasco}, {Dafonte}, {Diakit{\'e}},
  {Figueras}, {Garcia-Gutierrez}, {Garofalo}, {Helmi}, {Jim{\'e}nez-Arranz},
  {Jordi}, {Kervella}, {Kostrzewa-Rutkowska}, {Leclerc}, {Licata}, {Manteiga},
  {Masip}, {Mongui{\'o}}, {Ramos}, {Robichon}, {Robin}, {Romero-G{\'o}mez},
  {S{\'a}ez}, {Santove{\~n}a}, {Spina}, {Torralba Elipe}, \&
  {Weiler}}]{Babusiaux2022}
{Babusiaux}, C., {Fabricius}, C., {Khanna}, S., {et~al.} 2023, \aap, 674, A32

\bibitem[{{Babusiaux} \& {Gilmore}(2005)}]{Babusiaux2005}
{Babusiaux}, C. \& {Gilmore}, G. 2005, \mnras, 358, 1309

\bibitem[{{Barbuy} {et~al.}(2018){Barbuy}, {Chiappini}, \&
  {Gerhard}}]{Barbuy2018}
{Barbuy}, B., {Chiappini}, C., \& {Gerhard}, O. 2018, \araa, 56, 223

\bibitem[{{Battaglia} {et~al.}(2022){Battaglia}, {Taibi}, {Thomas}, \&
  {Fritz}}]{Battaglia2022}
{Battaglia}, G., {Taibi}, S., {Thomas}, G.~F., \& {Fritz}, T.~K. 2022, \aap,
  657, A54

\bibitem[{{Bellini} {et~al.}(2011){Bellini}, {Anderson}, \&
  {Bedin}}]{Bellini2011}
{Bellini}, A., {Anderson}, J., \& {Bedin}, L.~R. 2011, \pasp, 123, 622

\bibitem[{{Bellini} {et~al.}(2017){Bellini}, {Anderson}, {Bedin}, {King}, {van
  der Marel}, {Piotto}, \& {Cool}}]{Bellini_WcenCore2017}
{Bellini}, A., {Anderson}, J., {Bedin}, L.~R., {et~al.} 2017, \apj, 842, 6

\bibitem[{{Bellini} {et~al.}(2018){Bellini}, {Libralato}, {Bedin}, {Milone},
  {van der Marel}, {Anderson}, {Apai}, {Burgasser}, {Marino}, \&
  {Rees}}]{Bellini_WCENF12018}
{Bellini}, A., {Libralato}, M., {Bedin}, L.~R., {et~al.} 2018, \apj, 853, 86

\bibitem[{{Bernard} {et~al.}(2018){Bernard}, {Schultheis}, {Di Matteo}, {Hill},
  {Haywood}, \& {Calamida}}]{Bernard2018}
{Bernard}, E.~J., {Schultheis}, M., {Di Matteo}, P., {et~al.} 2018, \mnras,
  477, 3507

\bibitem[{{Brown} {et~al.}(2009){Brown}, {Sahu}, {Zoccali}, {Renzini},
  {Ferguson}, {Anderson}, {Smith}, {Bond}, {Minniti}, {Valenti}, {Casertano},
  {Livio}, {Panagia}, {Vanden Berg}, \& {Valenti}}]{BTS_Brown2009}
{Brown}, T.~M., {Sahu}, K., {Zoccali}, M., {et~al.} 2009, \aj, 137, 3172

\bibitem[{{Calamida} {et~al.}(2014){Calamida}, {Sahu}, {Anderson}, {Casertano},
  {Cassisi}, {Salaris}, {Brown}, {Sokol}, {Bond}, {Ferraro}, {Ferguson},
  {Livio}, {Valenti}, {Buonanno}, {Clarkson}, \&
  {Pietrinferni}}]{Calamida_SWEEPS2014}
{Calamida}, A., {Sahu}, K.~C., {Anderson}, J., {et~al.} 2014, \apj, 790, 164

\bibitem[{{Cantat-Gaudin} {et~al.}(2023){Cantat-Gaudin}, {Fouesneau}, {Rix},
  {Brown}, {Castro-Ginard}, {Kostrzewa-Rutkowska}, {Drimmel}, {Hogg}, {Casey},
  {Khanna}, {Oh}, {Price-Whelan}, {Belokurov}, {Saydjari}, \&
  {Green}}]{DR3SelFunct2022arXiv220809335C}
{Cantat-Gaudin}, T., {Fouesneau}, M., {Rix}, H.-W., {et~al.} 2023, \aap, 669,
  A55

\bibitem[{{Chatzopoulos} {et~al.}(2015){Chatzopoulos}, {Fritz}, {Gerhard},
  {Gillessen}, {Wegg}, {Genzel}, \& {Pfuhl}}]{Chatzopoulos2015}
{Chatzopoulos}, S., {Fritz}, T.~K., {Gerhard}, O., {et~al.} 2015, \mnras, 447,
  948

\bibitem[{{Clarke} \& {Gerhard}(2022)}]{Clarke2022}
{Clarke}, J.~P. \& {Gerhard}, O. 2022, \mnras, 512, 2171

\bibitem[{{Clarke} {et~al.}(2019){Clarke}, {Wegg}, {Gerhard}, {Smith}, {Lucas},
  \& {Wylie}}]{Clarke2019}
{Clarke}, J.~P., {Wegg}, C., {Gerhard}, O., {et~al.} 2019, \mnras, 489, 3519

\bibitem[{{Clarkson} {et~al.}(2008){Clarkson}, {Sahu}, {Anderson}, {Smith},
  {Brown}, {Rich}, {Casertano}, {Bond}, {Livio}, {Minniti}, {Panagia},
  {Renzini}, {Valenti}, \& {Zoccali}}]{Clarkson_BTS2008}
{Clarkson}, W., {Sahu}, K., {Anderson}, J., {et~al.} 2008, \apj, 684, 1110

\bibitem[{{Clarkson} {et~al.}(2018){Clarkson}, {Calamida}, {Sahu}, {Brown},
  {Gennaro}, {Avila}, {Valenti}, {Debattista}, {Rich}, {Minniti}, {Zoccali}, \&
  {Aufdemberge}}]{Clarkson2018}
{Clarkson}, W.~I., {Calamida}, A., {Sahu}, K.~C., {et~al.} 2018, \apj, 858, 46

\bibitem[{{del Pino} {et~al.}(2022){del Pino}, {Libralato}, {van der Marel},
  {Bennet}, {Fardal}, {Anderson}, {Bellini}, {Tony Sohn}, \&
  {Watkins}}]{delPino_GaiaHUB2022}
{del Pino}, A., {Libralato}, M., {van der Marel}, R.~P., {et~al.} 2022, \apj,
  933, 76

\bibitem[{{El-Badry} {et~al.}(2021){El-Badry}, {Rix}, \&
  {Heintz}}]{ElBadry2021}
{El-Badry}, K., {Rix}, H.-W., \& {Heintz}, T.~M. 2021, \mnras, 506, 2269

\bibitem[{{Everall} \& {Boubert}(2022)}]{EDR3SelFunct2022MNRAS.509.6205E}
{Everall}, A. \& {Boubert}, D. 2022, \mnras, 509, 6205

\bibitem[{{Fabricius} {et~al.}(2021){Fabricius}, {Luri}, {Arenou}, {Babusiaux},
  {Helmi}, {Muraveva}, {Reyl{\'e}}, {Spoto}, {Vallenari}, {Antoja}, {Balbinot},
  {Barache}, {Bauchet}, {Bragaglia}, {Busonero}, {Cantat-Gaudin}, {Carrasco},
  {Diakit{\'e}}, {Fabrizio}, {Figueras}, {Garcia-Gutierrez}, {Garofalo},
  {Jordi}, {Kervella}, {Khanna}, {Leclerc}, {Licata}, {Lambert}, {Marrese},
  {Masip}, {Ramos}, {Robichon}, {Robin}, {Romero-G{\'o}mez}, {Rubele}, \&
  {Weiler}}]{GaiaEDR3_Fabricius2021}
{Fabricius}, C., {Luri}, X., {Arenou}, F., {et~al.} 2021, \aap, 649, A5

\bibitem[{{Feldmeier-Krause} {et~al.}(2015){Feldmeier-Krause}, {Neumayer},
  {Sch{\"o}del}, {Seth}, {Hilker}, {de Zeeuw}, {Kuntschner}, {Walcher},
  {L{\"u}tzgendorf}, \& {Kissler-Patig}}]{FeldmeierKrause2015}
{Feldmeier-Krause}, A., {Neumayer}, N., {Sch{\"o}del}, R., {et~al.} 2015, \aap,
  584, A2

\bibitem[{{Fragkoudi} {et~al.}(2020){Fragkoudi}, {Grand}, {Pakmor},
  {Bl{\'a}zquez-Calero}, {Gargiulo}, {Gomez}, {Marinacci}, {Monachesi}, {Ness},
  {Perez}, {Tissera}, \& {White}}]{Fragkoudi2020}
{Fragkoudi}, F., {Grand}, R.~J.~J., {Pakmor}, R., {et~al.} 2020, \mnras, 494,
  5936

\bibitem[{{Gaia Collaboration} {et~al.}(2021){Gaia Collaboration}, {Brown},
  {Vallenari}, {Prusti}, {de Bruijne}, {Babusiaux}, {Biermann}, {Creevey},
  {Evans}, {Eyer}, {Hutton}, {Jansen}, {Jordi}, {Klioner}, {Lammers},
  {Lindegren}, {Luri}, {Mignard}, {Panem}, {Pourbaix}, {Randich}, {Sartoretti},
  {Soubiran}, {Walton}, {Arenou}, {Bailer-Jones}, {Bastian}, {Cropper},
  {Drimmel}, {Katz}, {Lattanzi}, {van Leeuwen}, {Bakker}, {Cacciari},
  {Casta{\~n}eda}, {De Angeli}, {Ducourant}, {Fabricius}, {Fouesneau},
  {Fr{\'e}mat}, {Guerra}, {Guerrier}, {Guiraud}, {Jean-Antoine Piccolo},
  {Masana}, {Messineo}, {Mowlavi}, {Nicolas}, {Nienartowicz}, {Pailler},
  {Panuzzo}, {Riclet}, {Roux}, {Seabroke}, {Sordo}, {Tanga}, {Th{\'e}venin},
  {Gracia-Abril}, {Portell}, {Teyssier}, {Altmann}, {Andrae}, {Bellas-Velidis},
  {Benson}, {Berthier}, {Blomme}, {Brugaletta}, {Burgess}, {Busso}, {Carry},
  {Cellino}, {Cheek}, {Clementini}, {Damerdji}, {Davidson}, {Delchambre},
  {Dell'Oro}, {Fern{\'a}ndez-Hern{\'a}ndez}, {Galluccio}, {Garc{\'\i}a-Lario},
  {Garcia-Reinaldos}, {Gonz{\'a}lez-N{\'u}{\~n}ez}, {Gosset}, {Haigron},
  {Halbwachs}, {Hambly}, {Harrison}, {Hatzidimitriou}, {Heiter},
  {Hern{\'a}ndez}, {Hestroffer}, {Hodgkin}, {Holl}, {Jan{\ss}en}, {Jevardat de
  Fombelle}, {Jordan}, {Krone-Martins}, {Lanzafame}, {L{\"o}ffler}, {Lorca},
  {Manteiga}, {Marchal}, {Marrese}, {Moitinho}, {Mora}, {Muinonen}, {Osborne},
  {Pancino}, {Pauwels}, {Petit}, {Recio-Blanco}, {Richards}, {Riello},
  {Rimoldini}, {Robin}, {Roegiers}, {Rybizki}, {Sarro}, {Siopis}, {Smith},
  {Sozzetti}, {Ulla}, {Utrilla}, {van Leeuwen}, {van Reeven}, {Abbas}, {Abreu
  Aramburu}, {Accart}, {Aerts}, {Aguado}, {Ajaj}, {Altavilla}, {{\'A}lvarez},
  {{\'A}lvarez Cid-Fuentes}, {Alves}, {Anderson}, {Anglada Varela}, {Antoja},
  {Audard}, {Baines}, {Baker}, {Balaguer-N{\'u}{\~n}ez}, {Balbinot}, {Balog},
  {Barache}, {Barbato}, {Barros}, {Barstow}, {Bartolom{\'e}}, {Bassilana},
  {Bauchet}, {Baudesson-Stella}, {Becciani}, {Bellazzini}, {Bernet}, {Bertone},
  {Bianchi}, {Blanco-Cuaresma}, {Boch}, {Bombrun}, {Bossini}, {Bouquillon},
  {Bragaglia}, {Bramante}, {Breedt}, {Bressan}, {Brouillet}, {Bucciarelli},
  {Burlacu}, {Busonero}, {Butkevich}, {Buzzi}, {Caffau}, {Cancelliere},
  {C{\'a}novas}, {Cantat-Gaudin}, {Carballo}, {Carlucci}, {Carnerero},
  {Carrasco}, {Casamiquela}, {Castellani}, {Castro-Ginard}, {Castro Sampol},
  {Chaoul}, {Charlot}, {Chemin}, {Chiavassa}, {Cioni}, {Comoretto}, {Cooper},
  {Cornez}, {Cowell}, {Crifo}, {Crosta}, {Crowley}, {Dafonte}, {Dapergolas},
  {David}, {David}, {de Laverny}, {De Luise}, {De March}, {De Ridder}, {de
  Souza}, {de Teodoro}, {de Torres}, {del Peloso}, {del Pozo}, {Delbo},
  {Delgado}, {Delgado}, {Delisle}, {Di Matteo}, {Diakite}, {Diener},
  {Distefano}, {Dolding}, {Eappachen}, {Edvardsson}, {Enke}, {Esquej}, {Fabre},
  {Fabrizio}, {Faigler}, {Fedorets}, {Fernique}, {Fienga}, {Figueras},
  {Fouron}, {Fragkoudi}, {Fraile}, {Franke}, {Gai}, {Garabato},
  {Garcia-Gutierrez}, {Garc{\'\i}a-Torres}, {Garofalo}, {Gavras}, {Gerlach},
  {Geyer}, {Giacobbe}, {Gilmore}, {Girona}, {Giuffrida}, {Gomel}, {Gomez},
  {Gonzalez-Santamaria}, {Gonz{\'a}lez-Vidal}, {Granvik},
  {Guti{\'e}rrez-S{\'a}nchez}, {Guy}, {Hauser}, {Haywood}, {Helmi}, {Hidalgo},
  {Hilger}, {H{\l}adczuk}, {Hobbs}, {Holland}, {Huckle}, {Jasniewicz},
  {Jonker}, {Juaristi Campillo}, {Julbe}, {Karbevska}, {Kervella}, {Khanna},
  {Kochoska}, {Kontizas}, {Kordopatis}, {Korn}, {Kostrzewa-Rutkowska},
  {Kruszy{\'n}ska}, {Lambert}, {Lanza}, {Lasne}, {Le Campion}, {Le Fustec},
  {Lebreton}, {Lebzelter}, {Leccia}, {Leclerc}, {Lecoeur-Taibi}, {Liao},
  {Licata}, {Lindstr{\o}m}, {Lister}, {Livanou}, {Lobel}, {Madrero Pardo},
  {Managau}, {Mann}, {Marchant}, {Marconi}, {Marcos Santos}, {Marinoni},
  {Marocco}, {Marshall}, {Martin Polo}, {Mart{\'\i}n-Fleitas}, {Masip},
  {Massari}, {Mastrobuono-Battisti}, {Mazeh}, {McMillan}, {Messina},
  {Michalik}, {Millar}, {Mints}, {Molina}, {Molinaro}, {Moln{\'a}r},
  {Montegriffo}, {Mor}, {Morbidelli}, {Morel}, {Morris}, {Mulone}, {Munoz},
  {Muraveva}, {Murphy}, {Musella}, {Noval}, {Ord{\'e}novic}, {Orr{\`u}},
  {Osinde}, {Pagani}, {Pagano}, {Palaversa}, {Palicio}, {Panahi}, {Pawlak},
  {Pe{\~n}alosa Esteller}, {Penttil{\"a}}, {Piersimoni}, {Pineau}, {Plachy},
  {Plum}, {Poggio}, {Poretti}, {Poujoulet}, {Pr{\v{s}}a}, {Pulone}, {Racero},
  {Ragaini}, {Rainer}, {Raiteri}, {Rambaux}, {Ramos}, {Ramos-Lerate}, {Re
  Fiorentin}, {Regibo}, {Reyl{\'e}}, {Ripepi}, {Riva}, {Rixon}, {Robichon},
  {Robin}, {Roelens}, {Rohrbasser}, {Romero-G{\'o}mez}, {Rowell}, {Royer},
  {Rybicki}, {Sadowski}, {Sagrist{\`a} Sell{\'e}s}, {Sahlmann}, {Salgado},
  {Salguero}, {Samaras}, {Sanchez Gimenez}, {Sanna}, {Santove{\~n}a},
  {Sarasso}, {Schultheis}, {Sciacca}, {Segol}, {Segovia}, {S{\'e}gransan},
  {Semeux}, {Shahaf}, {Siddiqui}, {Siebert}, {Siltala}, {Slezak}, {Smart},
  {Solano}, {Solitro}, {Souami}, {Souchay}, {Spagna}, {Spoto}, {Steele},
  {Steidelm{\"u}ller}, {Stephenson}, {S{\"u}veges}, {Szabados}, {Szegedi-Elek},
  {Taris}, {Tauran}, {Taylor}, {Teixeira}, {Thuillot}, {Tonello}, {Torra},
  {Torra}, {Turon}, {Unger}, {Vaillant}, {van Dillen}, {Vanel}, {Vecchiato},
  {Viala}, {Vicente}, {Voutsinas}, {Weiler}, {Wevers}, {Wyrzykowski}, {Yoldas},
  {Yvard}, {Zhao}, {Zorec}, {Zucker}, {Zurbach}, \& {Zwitter}}]{Gaia_BrownEDR3}
{Gaia Collaboration}, {Brown}, A.~G.~A., {Vallenari}, A., {et~al.} 2021, \aap,
  649, A1

\bibitem[{{Gaia Collaboration} {et~al.}(2022){Gaia Collaboration}, {Vallenari,
  A.}, {Brown, A.G.A.}, {Prusti, T.}, \& {et al.}}]{GaiaDR3}
{Gaia Collaboration}, {Vallenari, A.}, {Brown, A.G.A.}, {Prusti, T.}, \& {et
  al.} 2022, A\&A

\bibitem[{{Garro} {et~al.}(2022){Garro}, {Minniti}, {G{\'o}mez},
  {Fern{\'a}ndez-Trincado}, {Alonso-Garc{\'\i}a}, {Hempel}, \& {Zelada
  Bacigalupo}}]{Garro2022}
{Garro}, E.~R., {Minniti}, D., {G{\'o}mez}, M., {et~al.} 2022, \aap, 662, A95

\bibitem[{{Gillessen} {et~al.}(2009){Gillessen}, {Eisenhauer}, {Fritz},
  {Bartko}, {Dodds-Eden}, {Pfuhl}, {Ott}, \& {Genzel}}]{gillessen2009}
{Gillessen}, S., {Eisenhauer}, F., {Fritz}, T.~K., {et~al.} 2009, \apjl, 707,
  L114

\bibitem[{{Gonzalez} {et~al.}(2013){Gonzalez}, {Rejkuba}, {Zoccali}, {Valent},
  {Minniti}, \& {Tobar}}]{gonzalez2013}
{Gonzalez}, O.~A., {Rejkuba}, M., {Zoccali}, M., {et~al.} 2013, \aap, 552, A110

\bibitem[{{Gonzalez} {et~al.}(2012){Gonzalez}, {Rejkuba}, {Zoccali}, {Valenti},
  {Minniti}, {Schultheis}, {Tobar}, \& {Chen}}]{gonzalez2012}
{Gonzalez}, O.~A., {Rejkuba}, M., {Zoccali}, M., {et~al.} 2012, \aap, 543, A13

\bibitem[{{Gonzalez} {et~al.}(2015){Gonzalez}, {Zoccali}, {Debattista},
  {Alonso-Garc{\'\i}a}, {Valenti}, \& {Minniti}}]{gonzalez2015}
{Gonzalez}, O.~A., {Zoccali}, M., {Debattista}, V.~P., {et~al.} 2015, \aap,
  583, L5

\bibitem[{{GRAVITY Collaboration} {et~al.}(2018){GRAVITY Collaboration},
  {Abuter}, {Amorim}, {Anugu}, {Baub{\"o}ck}, {Benisty}, {Berger}, {Blind},
  {Bonnet}, {Brandner}, {Buron}, {Collin}, {Chapron}, {Cl{\'e}net}, {Coud{\'e}
  Du Foresto}, {de Zeeuw}, {Deen}, {Delplancke-Str{\"o}bele}, {Dembet},
  {Dexter}, {Duvert}, {Eckart}, {Eisenhauer}, {Finger}, {F{\"o}rster
  Schreiber}, {F{\'e}dou}, {Garcia}, {Garcia Lopez}, {Gao}, {Gendron},
  {Genzel}, {Gillessen}, {Gordo}, {Habibi}, {Haubois}, {Haug}, {Hau{\ss}mann},
  {Henning}, {Hippler}, {Horrobin}, {Hubert}, {Hubin}, {Jimenez Rosales},
  {Jochum}, {Jocou}, {Kaufer}, {Kellner}, {Kendrew}, {Kervella}, {Kok},
  {Kulas}, {Lacour}, {Lapeyr{\`e}re}, {Lazareff}, {Le Bouquin}, {L{\'e}na},
  {Lippa}, {Lenzen}, {M{\'e}rand}, {M{\"u}ler}, {Neumann}, {Ott}, {Palanca},
  {Paumard}, {Pasquini}, {Perraut}, {Perrin}, {Pfuhl}, {Plewa}, {Rabien},
  {Ram{\'\i}rez}, {Ramos}, {Rau}, {Rodr{\'\i}guez-Coira}, {Rohloff}, {Rousset},
  {Sanchez-Bermudez}, {Scheithauer}, {Sch{\"o}ller}, {Schuler}, {Spyromilio},
  {Straub}, {Straubmeier}, {Sturm}, {Tacconi}, {Tristram}, {Vincent}, {von
  Fellenberg}, {Wank}, {Waisberg}, {Widmann}, {Wieprecht}, {Wiest},
  {Wiezorrek}, {Woillez}, {Yazici}, {Ziegler}, \& {Zins}}]{GRAVITY2018}
{GRAVITY Collaboration}, {Abuter}, R., {Amorim}, A., {et~al.} 2018, \aap, 615,
  L15

\bibitem[{{GRAVITY Collaboration} {et~al.}(2020){GRAVITY Collaboration},
  {Abuter}, {Amorim}, {Baub{\"o}ck}, {Berger}, {Bonnet}, {Brandner}, {Cardoso},
  {Cl{\'e}net}, {de Zeeuw}, {Dexter}, {Eckart}, {Eisenhauer}, {F{\"o}rster
  Schreiber}, {Garcia}, {Gao}, {Gendron}, {Genzel}, {Gillessen}, {Habibi},
  {Haubois}, {Henning}, {Hippler}, {Horrobin}, {Jim{\'e}nez-Rosales}, {Jochum},
  {Jocou}, {Kaufer}, {Kervella}, {Lacour}, {Lapeyr{\`e}re}, {Le Bouquin},
  {L{\'e}na}, {Nowak}, {Ott}, {Paumard}, {Perraut}, {Perrin}, {Pfuhl},
  {Rodr{\'\i}guez-Coira}, {Shangguan}, {Scheithauer}, {Stadler}, {Straub},
  {Straubmeier}, {Sturm}, {Tacconi}, {Vincent}, {von Fellenberg}, {Waisberg},
  {Widmann}, {Wieprecht}, {Wiezorrek}, {Woillez}, {Yazici}, \&
  {Zins}}]{GRAVITY2020}
{GRAVITY Collaboration}, {Abuter}, R., {Amorim}, A., {et~al.} 2020, \aap, 636,
  L5

\bibitem[{Harris {et~al.}(2020)Harris, Millman, van~der Walt, Gommers,
  Virtanen, Cournapeau, Wieser, Taylor, Berg, Smith, Kern, Picus, Hoyer, van
  Kerkwijk, Brett, Haldane, del R{'{\i}}o, Wiebe, Peterson,
  G{'{e}}rard-Marchant, Sheppard, Reddy, Weckesser, Abbasi, Gohlke, \&
  Oliphant}]{harris2020array}
Harris, C.~R., Millman, K.~J., van~der Walt, S.~J., {et~al.} 2020, Nature, 585,
  357

\bibitem[{{Horta} {et~al.}(2021){Horta}, {Schiavon}, {Mackereth}, {Pfeffer},
  {Mason}, {Kisku}, {Fragkoudi}, {Allende Prieto}, {Cunha}, {Hasselquist},
  {Holtzman}, {Majewski}, {Nataf}, {O'Connell}, {Schultheis}, \&
  {Smith}}]{Horta2021}
{Horta}, D., {Schiavon}, R.~P., {Mackereth}, J.~T., {et~al.} 2021, \mnras, 500,
  1385

\bibitem[{{Johnson} {et~al.}(2011){Johnson}, {Rich}, {Fulbright}, {Valenti}, \&
  {McWilliam}}]{Johnson2011_PlautWindow}
{Johnson}, C.~I., {Rich}, R.~M., {Fulbright}, J.~P., {Valenti}, E., \&
  {McWilliam}, A. 2011, \apj, 732, 108

\bibitem[{{Kader} {et~al.}(2022){Kader}, {Pilachowski}, {Johnson}, {Rich},
  {Young}, {Simion}, {Clarkson}, {Michael}, {Kunder}, {Vivas}, {Koch-Hansen},
  \& {Marchetti}}]{Kader2022}
{Kader}, J.~A., {Pilachowski}, C.~A., {Johnson}, C.~I., {et~al.} 2022, \apj,
  940, 76

\bibitem[{{Kozhurina-Platais} \& {Martlin}(2021)}]{KozhurinaP2021}
{Kozhurina-Platais}, V. \& {Martlin}, C. 2021, {Accuracy of the HST/WFC3
  Standard Astrometric Catalog w.r.t Gaia EDR3}, Instrument Science Report WFC3
  2021-07, 18 pages

\bibitem[{{Kunder} {et~al.}(2016){Kunder}, {Rich}, {Koch}, {Storm}, {Nataf},
  {De Propris}, {Walker}, {Bono}, {Johnson}, {Shen}, \& {Li}}]{Kunder2016}
{Kunder}, A., {Rich}, R.~M., {Koch}, A., {et~al.} 2016, \apjl, 821, L25

\bibitem[{{Libralato} {et~al.}(2018){Libralato}, {Bellini}, {Bedin}, {Moreno
  D.}, {Fern{\'a}ndez-Trincado}, {Pichardo}, {van der Marel}, {Anderson},
  {Apai}, {Burgasser}, {Fabiola Marino}, {Milone}, {Rees}, \&
  {Watkins}}]{Libralato_F12018}
{Libralato}, M., {Bellini}, A., {Bedin}, L.~R., {et~al.} 2018, \apj, 854, 45

\bibitem[{{Libralato} {et~al.}(2022){Libralato}, {Bellini}, {Vesperini},
  {Piotto}, {Milone}, {van der Marel}, {Anderson}, {Aparicio}, {Barbuy},
  {Bedin}, {Borsato}, {Cassisi}, {Dalessandro}, {Ferraro}, {King}, {Lanzoni},
  {Nardiello}, {Ortolani}, {Sarajedini}, \& {Sohn}}]{Libralato_GCHST2022}
{Libralato}, M., {Bellini}, A., {Vesperini}, E., {et~al.} 2022, \apj, 934, 150

\bibitem[{{Lindegren} {et~al.}(2021{\natexlab{a}}){Lindegren}, {Bastian},
  {Biermann}, {Bombrun}, {de Torres}, {Gerlach}, {Geyer}, {Hern{\'a}ndez},
  {Hilger}, {Hobbs}, {Klioner}, {Lammers}, {McMillan}, {Ramos-Lerate},
  {Steidelm{\"u}ller}, {Stephenson}, \& {van Leeuwen}}]{Lindegren2021}
{Lindegren}, L., {Bastian}, U., {Biermann}, M., {et~al.} 2021{\natexlab{a}},
  \aap, 649, A4

\bibitem[{{Lindegren} {et~al.}(2018){Lindegren}, {Hern{\'a}ndez}, {Bombrun},
  {Klioner}, {Bastian}, {Ramos-Lerate}, {de Torres}, {Steidelm{\"u}ller},
  {Stephenson}, {Hobbs}, {Lammers}, {Biermann}, {Geyer}, {Hilger}, {Michalik},
  {Stampa}, {McMillan}, {Casta{\~n}eda}, {Clotet}, {Comoretto}, {Davidson},
  {Fabricius}, {Gracia}, {Hambly}, {Hutton}, {Mora}, {Portell}, {van Leeuwen},
  {Abbas}, {Abreu}, {Altmann}, {Andrei}, {Anglada}, {Balaguer-N{\'u}{\~n}ez},
  {Barache}, {Becciani}, {Bertone}, {Bianchi}, {Bouquillon}, {Bourda},
  {Br{\"u}semeister}, {Bucciarelli}, {Busonero}, {Buzzi}, {Cancelliere},
  {Carlucci}, {Charlot}, {Cheek}, {Crosta}, {Crowley}, {de Bruijne}, {de
  Felice}, {Drimmel}, {Esquej}, {Fienga}, {Fraile}, {Gai}, {Garralda},
  {Gonz{\'a}lez-Vidal}, {Guerra}, {Hauser}, {Hofmann}, {Holl}, {Jordan},
  {Lattanzi}, {Lenhardt}, {Liao}, {Licata}, {Lister}, {L{\"o}ffler},
  {Marchant}, {Martin-Fleitas}, {Messineo}, {Mignard}, {Morbidelli}, {Poggio},
  {Riva}, {Rowell}, {Salguero}, {Sarasso}, {Sciacca}, {Siddiqui}, {Smart},
  {Spagna}, {Steele}, {Taris}, {Torra}, {van Elteren}, {van Reeven}, \&
  {Vecchiato}}]{Lindegren2018}
{Lindegren}, L., {Hern{\'a}ndez}, J., {Bombrun}, A., {et~al.} 2018, \aap, 616,
  A2

\bibitem[{{Lindegren} {et~al.}(2021{\natexlab{b}}){Lindegren}, {Klioner},
  {Hern{\'a}ndez}, {Bombrun}, {Ramos-Lerate}, {Steidelm{\"u}ller}, {Bastian},
  {Biermann}, {de Torres}, {Gerlach}, {Geyer}, {Hilger}, {Hobbs}, {Lammers},
  {McMillan}, {Stephenson}, {Casta{\~n}eda}, {Davidson}, {Fabricius},
  {Gracia-Abril}, {Portell}, {Rowell}, {Teyssier}, {Torra}, {Bartolom{\'e}},
  {Clotet}, {Garralda}, {Gonz{\'a}lez-Vidal}, {Torra}, {Abbas}, {Altmann},
  {Anglada Varela}, {Balaguer-N{\'u}{\~n}ez}, {Balog}, {Barache}, {Becciani},
  {Bernet}, {Bertone}, {Bianchi}, {Bouquillon}, {Brown}, {Bucciarelli},
  {Busonero}, {Butkevich}, {Buzzi}, {Cancelliere}, {Carlucci}, {Charlot},
  {Cioni}, {Crosta}, {Crowley}, {del Peloso}, {del Pozo}, {Drimmel}, {Esquej},
  {Fienga}, {Fraile}, {Gai}, {Garcia-Reinaldos}, {Guerra}, {Hambly}, {Hauser},
  {Jan{\ss}en}, {Jordan}, {Kostrzewa-Rutkowska}, {Lattanzi}, {Liao}, {Licata},
  {Lister}, {L{\"o}ffler}, {Marchant}, {Masip}, {Mignard}, {Mints}, {Molina},
  {Mora}, {Morbidelli}, {Murphy}, {Pagani}, {Panuzzo}, {Pe{\~n}alosa Esteller},
  {Poggio}, {Re Fiorentin}, {Riva}, {Sagrist{\`a} Sell{\'e}s}, {Sanchez
  Gimenez}, {Sarasso}, {Sciacca}, {Siddiqui}, {Smart}, {Souami}, {Spagna},
  {Steele}, {Taris}, {Utrilla}, {van Reeven}, \&
  {Vecchiato}}]{2021A&A...649A...2L}
{Lindegren}, L., {Klioner}, S.~A., {Hern{\'a}ndez}, J., {et~al.}
  2021{\natexlab{b}}, \aap, 649, A2

\bibitem[{{Lindegren} {et~al.}(2016){Lindegren}, {Lammers}, {Bastian},
  {Hern{\'a}ndez}, {Klioner}, {Hobbs}, {Bombrun}, {Michalik}, {Ramos-Lerate},
  {Butkevich}, {Comoretto}, {Joliet}, {Holl}, {Hutton}, {Parsons},
  {Steidelm{\"u}ller}, {Abbas}, {Altmann}, {Andrei}, {Anton}, {Bach},
  {Barache}, {Becciani}, {Berthier}, {Bianchi}, {Biermann}, {Bouquillon},
  {Bourda}, {Br{\"u}semeister}, {Bucciarelli}, {Busonero}, {Carlucci},
  {Casta{\~n}eda}, {Charlot}, {Clotet}, {Crosta}, {Davidson}, {de Felice},
  {Drimmel}, {Fabricius}, {Fienga}, {Figueras}, {Fraile}, {Gai}, {Garralda},
  {Geyer}, {Gonz{\'a}lez-Vidal}, {Guerra}, {Hambly}, {Hauser}, {Jordan},
  {Lattanzi}, {Lenhardt}, {Liao}, {L{\"o}ffler}, {McMillan}, {Mignard}, {Mora},
  {Morbidelli}, {Portell}, {Riva}, {Sarasso}, {Serraller}, {Siddiqui}, {Smart},
  {Spagna}, {Stampa}, {Steele}, {Taris}, {Torra}, {van Reeven}, {Vecchiato},
  {Zschocke}, {de Bruijne}, {Gracia}, {Raison}, {Lister}, {Marchant},
  {Messineo}, {Soffel}, {Osorio}, {de Torres}, \& {O'Mullane}}]{Lindegren2016}
{Lindegren}, L., {Lammers}, U., {Bastian}, U., {et~al.} 2016, \aap, 595, A4

\bibitem[{{Lucey} {et~al.}(2021){Lucey}, {Hawkins}, {Ness}, {Debattista},
  {Luna}, {Asplund}, {Bensby}, {Casagrande}, {Feltzing}, {Freeman},
  {Kobayashi}, \& {Marino}}]{Lucey2021}
{Lucey}, M., {Hawkins}, K., {Ness}, M., {et~al.} 2021, \mnras, 501, 5981

\bibitem[{{Luna} {et~al.}(2019){Luna}, {Minniti}, \&
  {Alonso-Garc{\'\i}a}}]{Luna2019}
{Luna}, A., {Minniti}, D., \& {Alonso-Garc{\'\i}a}, J. 2019, \apjl, 887, L39

\bibitem[{{Ma{\'\i}z Apell{\'a}niz} {et~al.}(2021){Ma{\'\i}z Apell{\'a}niz},
  {Pantaleoni Gonz{\'a}lez}, \& {Barb{\'a}}}]{MaizApellaniz2021}
{Ma{\'\i}z Apell{\'a}niz}, J., {Pantaleoni Gonz{\'a}lez}, M., \& {Barb{\'a}},
  R.~H. 2021, \aap, 649, A13

\bibitem[{{Marchetti} {et~al.}(2022){Marchetti}, {Johnson}, {Joyce}, {Rich},
  {Simion}, {Young}, {Clarkson}, {Pilachowski}, {Michael}, {Kunder}, \&
  {Koch-Hansen}}]{Marchetti2022}
{Marchetti}, T., {Johnson}, C.~I., {Joyce}, M., {et~al.} 2022, \aap, 664, A124

\bibitem[{{Massari} {et~al.}(2018){Massari}, {Breddels}, {Helmi}, {Posti},
  {Brown}, \& {Tolstoy}}]{Massari2018}
{Massari}, D., {Breddels}, M.~A., {Helmi}, A., {et~al.} 2018, Nature Astronomy,
  2, 156

\bibitem[{{Massari} {et~al.}(2020){Massari}, {Helmi}, {Mucciarelli}, {Sales},
  {Spina}, \& {Tolstoy}}]{Massari2020}
{Massari}, D., {Helmi}, A., {Mucciarelli}, A., {et~al.} 2020, \aap, 633, A36

\bibitem[{{McWilliam} \& {Zoccali}(2010)}]{McWilliam+Zoccali2010}
{McWilliam}, A. \& {Zoccali}, M. 2010, \apj, 724, 1491

\bibitem[{{Minniti} {et~al.}(2010){Minniti}, {Lucas}, {Emerson}, {Saito},
  {Hempel}, {Pietrukowicz}, {Ahumada}, {Alonso}, {Alonso-Garcia}, {Arias},
  {Bandyopadhyay}, {Barb{\'a}}, {Barbuy}, {Bedin}, {Bica}, {Borissova},
  {Bronfman}, {Carraro}, {Catelan}, {Clari{\'a}}, {Cross}, {de Grijs},
  {D{\'e}k{\'a}ny}, {Drew}, {Fari{\~n}a}, {Feinstein}, {Fern{\'a}ndez
  Laj{\'u}s}, {Gamen}, {Geisler}, {Gieren}, {Goldman}, {Gonzalez}, {Gunthardt},
  {Gurovich}, {Hambly}, {Irwin}, {Ivanov}, {Jord{\'a}n}, {Kerins}, {Kinemuchi},
  {Kurtev}, {L{\'o}pez-Corredoira}, {Maccarone}, {Masetti}, {Merlo},
  {Messineo}, {Mirabel}, {Monaco}, {Morelli}, {Padilla}, {Palma}, {Parisi},
  {Pignata}, {Rejkuba}, {Roman-Lopes}, {Sale}, {Schreiber}, {Schr{\"o}der},
  {Smith}, {}, {Soto}, {Tamura}, {Tappert}, {Thompson}, {Toledo}, {Zoccali}, \&
  {Pietrzynski}}]{Minniti2010}
{Minniti}, D., {Lucas}, P.~W., {Emerson}, J.~P., {et~al.} 2010, \na, 15, 433

\bibitem[{{Nataf} {et~al.}(2016){Nataf}, {Gonzalez}, {Casagrande}, {Zasowski},
  {Wegg}, {Wolf}, {Kunder}, {Alonso-Garcia}, {Minniti}, {Rejkuba}, {Saito},
  {Valenti}, {Zoccali}, {Poleski}, {Pietrzy{\'n}ski}, {Skowron},
  {Soszy{\'n}ski}, {Szyma{\'n}ski}, {Udalski}, {Ulaczyk}, \&
  {Wyrzykowski}}]{Nataf2016}
{Nataf}, D.~M., {Gonzalez}, O.~A., {Casagrande}, L., {et~al.} 2016, \mnras,
  456, 2692

\bibitem[{{Nataf} {et~al.}(2010){Nataf}, {Udalski}, {Gould}, {Fouqu{\'e}}, \&
  {Stanek}}]{Nataf2010}
{Nataf}, D.~M., {Udalski}, A., {Gould}, A., {Fouqu{\'e}}, P., \& {Stanek},
  K.~Z. 2010, \apjl, 721, L28

\bibitem[{{Ness} \& {Lang}(2016)}]{Ness2016}
{Ness}, M. \& {Lang}, D. 2016, \aj, 152, 14

\bibitem[{{Nidever} {et~al.}(2012){Nidever}, {Zasowski}, \&
  {Majewski}}]{nidever2012}
{Nidever}, D.~L., {Zasowski}, G., \& {Majewski}, S.~R. 2012, \apjs, 201, 35

\bibitem[{{Nogueras-Lara}(2022)}]{NoguerasLara2022}
{Nogueras-Lara}, F. 2022, \aap, 666, A72

\bibitem[{{Nogueras-Lara} {et~al.}(2020){Nogueras-Lara}, {Sch{\"o}del},
  {Gallego-Calvente}, {Gallego-Cano}, {Shahzamanian}, {Dong}, {Neumayer},
  {Hilker}, {Najarro}, {Nishiyama}, {Feldmeier-Krause}, {Girard}, \&
  {Cassisi}}]{Nogueras-Lara2020}
{Nogueras-Lara}, F., {Sch{\"o}del}, R., {Gallego-Calvente}, A.~T., {et~al.}
  2020, Nature Astronomy, 4, 377

\bibitem[{{Nogueras-Lara} {et~al.}(2021){Nogueras-Lara}, {Sch{\"o}del}, \&
  {Neumayer}}]{Nogueras-Lara2021}
{Nogueras-Lara}, F., {Sch{\"o}del}, R., \& {Neumayer}, N. 2021, \aap, 653, A133

\bibitem[{Pedregosa {et~al.}(2011)Pedregosa, Varoquaux, Gramfort, Michel,
  Thirion, Grisel, Blondel, Prettenhofer, Weiss, Dubourg, Vanderplas, Passos,
  Cournapeau, Brucher, Perrot, \& Duchesnay}]{scikit-learn}
Pedregosa, F., Varoquaux, G., Gramfort, A., {et~al.} 2011, Journal of Machine
  Learning Research, 12, 2825

\bibitem[{{Pfuhl} {et~al.}(2011){Pfuhl}, {Fritz}, {Zilka}, {Maness},
  {Eisenhauer}, {Genzel}, {Gillessen}, {Ott}, {Dodds-Eden}, \&
  {Sternberg}}]{pfuhl2011}
{Pfuhl}, O., {Fritz}, T.~K., {Zilka}, M., {et~al.} 2011, \apj, 741, 108

\bibitem[{{Portail} {et~al.}(2017){Portail}, {Gerhard}, {Wegg}, \&
  {Ness}}]{portail2017}
{Portail}, M., {Gerhard}, O., {Wegg}, C., \& {Ness}, M. 2017, \mnras, 465, 1621

\bibitem[{{Riello} {et~al.}(2021){Riello}, {De Angeli}, {Evans}, {Montegriffo},
  {Carrasco}, {Busso}, {Palaversa}, {Burgess}, {Diener}, {Davidson}, {Rowell},
  {Fabricius}, {Jordi}, {Bellazzini}, {Pancino}, {Harrison}, {Cacciari}, {van
  Leeuwen}, {Hambly}, {Hodgkin}, {Osborne}, {Altavilla}, {Barstow}, {Brown},
  {Castellani}, {Cowell}, {De Luise}, {Gilmore}, {Giuffrida}, {Hidalgo},
  {Holland}, {Marinoni}, {Pagani}, {Piersimoni}, {Pulone}, {Ragaini}, {Rainer},
  {Richards}, {Sanna}, {Walton}, {Weiler}, \& {Yoldas}}]{Riello2021}
{Riello}, M., {De Angeli}, F., {Evans}, D.~W., {et~al.} 2021, \aap, 649, A3

\bibitem[{{Rix} {et~al.}(2022){Rix}, {Chandra}, {Andrae}, {Price-Whelan},
  {Weinberg}, {Conroy}, {Fouesneau}, {Hogg}, {De Angeli}, {Naidu}, {Xiang}, \&
  {Ruz-Mieres}}]{Rix2022}
{Rix}, H.-W., {Chandra}, V., {Andrae}, R., {et~al.} 2022, \apj, 941, 45

\bibitem[{{Rybizki} {et~al.}(2022){Rybizki}, {Green}, {Rix}, {El-Badry},
  {Demleitner}, {Zari}, {Udalski}, {Smart}, \& {Gould}}]{Rybizki2022MNRAS}
{Rybizki}, J., {Green}, G.~M., {Rix}, H.-W., {et~al.} 2022, \mnras, 510, 2597

\bibitem[{{Sahu} {et~al.}(2006){Sahu}, {Casertano}, {Bond}, {Valenti}, {Ed
  Smith}, {Minniti}, {Zoccali}, {Livio}, {Panagia}, {Piskunov}, {Brown},
  {Brown}, {Renzini}, {Rich}, {Clarkson}, \& {Lubow}}]{Sahu2006}
{Sahu}, K.~C., {Casertano}, S., {Bond}, H.~E., {et~al.} 2006, \nat, 443, 534

\bibitem[{{Saito} {et~al.}(2012){Saito}, {Hempel}, {Minniti}, {Lucas},
  {Rejkuba}, {Toledo}, {Gonzalez}, {Alonso-Garc{\'\i}a}, {Irwin},
  {Gonzalez-Solares}, {Hodgkin}, {Lewis}, {Cross}, {Ivanov}, {Kerins},
  {Emerson}, {Soto}, {Am{\^o}res}, {Gurovich}, {D{\'e}k{\'a}ny}, {Angeloni},
  {Beamin}, {Catelan}, {Padilla}, {Zoccali}, {Pietrukowicz}, {Moni Bidin},
  {Mauro}, {Geisler}, {Folkes}, {Sale}, {Borissova}, {Kurtev}, {Ahumada},
  {Alonso}, {Adamson}, {Arias}, {Bandyopadhyay}, {Barb{\'a}}, {Barbuy},
  {Baume}, {Bedin}, {Bellini}, {Benjamin}, {Bica}, {Bonatto}, {Bronfman},
  {Carraro}, {Chen{\`e}}, {Clari{\'a}}, {Clarke}, {Contreras}, {Corvill{\'o}n},
  {de Grijs}, {Dias}, {Drew}, {Fari{\~n}a}, {Feinstein},
  {Fern{\'a}ndez-Laj{\'u}s}, {Gamen}, {Gieren}, {Goldman},
  {Gonz{\'a}lez-Fern{\'a}ndez}, {Grand}, {Gunthardt}, {Hambly}, {Hanson},
  {He{\l}miniak}, {Hoare}, {Huckvale}, {Jord{\'a}n}, {Kinemuchi}, {Longmore},
  {L{\'o}pez-Corredoira}, {Maccarone}, {Majaess}, {Mart{\'\i}n}, {Masetti},
  {Mennickent}, {Mirabel}, {Monaco}, {Morelli}, {Motta}, {Palma}, {Parisi},
  {Parker}, {Pe{\~n}aloza}, {Pietrzy{\'n}ski}, {Pignata}, {Popescu}, {Read},
  {Rojas}, {Roman-Lopes}, {Ruiz}, {Saviane}, {Schreiber}, {Schr{\"o}der},
  {Sharma}, {Smith}, {Sodr{\'e}}, {Stead}, {Stephens}, {Tamura}, {Tappert},
  {Thompson}, {Valenti}, {Vanzi}, {Walton}, {Weidmann}, \&
  {Zijlstra}}]{Saito2012}
{Saito}, R.~K., {Hempel}, M., {Minniti}, D., {et~al.} 2012, \aap, 537, A107

\bibitem[{{Sanders} {et~al.}(2022){Sanders}, {Smith},
  {Gonz{\'a}lez-Fern{\'a}ndez}, {Lucas}, \& {Minniti}}]{Sanders2022}
{Sanders}, J.~L., {Smith}, L., {Gonz{\'a}lez-Fern{\'a}ndez}, C., {Lucas}, P.,
  \& {Minniti}, D. 2022, \mnras, 514, 2407

\bibitem[{{Scalco} {et~al.}(2021){Scalco}, {Bellini}, {Bedin}, {Anderson},
  {Rosati}, {Libralato}, {Salaris}, {Vesperini}, {Nardiello}, {Apai},
  {Burgasser}, \& {Gerasimov}}]{Scalco_WCENF2F32021}
{Scalco}, M., {Bellini}, A., {Bedin}, L.~R., {et~al.} 2021, \mnras, 505, 3549

\bibitem[{{Sch{\"o}del} {et~al.}(2014){Sch{\"o}del}, {Feldmeier}, {Neumayer},
  {Meyer}, \& {Yelda}}]{schoedel2014}
{Sch{\"o}del}, R., {Feldmeier}, A., {Neumayer}, N., {Meyer}, L., \& {Yelda}, S.
  2014, Classical and Quantum Gravity, 31, 244007

\bibitem[{{Schultheis} {et~al.}(2014){Schultheis}, {Chen}, {Jiang}, {Gonzalez},
  {Enokiya}, {Fukui}, {Torii}, {Rejkuba}, \& {Minniti}}]{schultheis2014}
{Schultheis}, M., {Chen}, B.~Q., {Jiang}, B.~W., {et~al.} 2014, \aap, 566, A120

\bibitem[{{Shahzamanian} {et~al.}(2022){Shahzamanian}, {Sch{\"o}del},
  {Nogueras-Lara}, {Mart{\'\i}nez-Arranz}, {Sormani}, {Gallego-Calvente},
  {Gallego-Cano}, \& {Alburai}}]{Shahzamanian2022}
{Shahzamanian}, B., {Sch{\"o}del}, R., {Nogueras-Lara}, F., {et~al.} 2022,
  \aap, 662, A11

\bibitem[{{Simion} {et~al.}(2017){Simion}, {Belokurov}, {Irwin}, {Koposov},
  {Gonzalez-Fernandez}, {Robin}, {Shen}, \& {Li}}]{simion2017}
{Simion}, I.~T., {Belokurov}, V., {Irwin}, M., {et~al.} 2017, \mnras, 471, 4323

\bibitem[{{Smith} {et~al.}(2018){Smith}, {Lucas}, {Kurtev}, {Smart}, {Minniti},
  {Borissova}, {Jones}, {Zhang}, {Marocco}, {Contreras Pe{\~n}a}, {Gromadzki},
  {Kuhn}, {Drew}, {Pinfield}, \& {Bedin}}]{Smith2018}
{Smith}, L.~C., {Lucas}, P.~W., {Kurtev}, R., {et~al.} 2018, \mnras, 474, 1826

\bibitem[{{Stanek} {et~al.}(1994){Stanek}, {Mateo}, {Udalski}, {Szymanski},
  {Kaluzny}, \& {Kubiak}}]{Stanek1994}
{Stanek}, K.~Z., {Mateo}, M., {Udalski}, A., {et~al.} 1994, \apjl, 429, L73

\bibitem[{{Surot} {et~al.}(2020){Surot}, {Valenti}, {Gonzalez}, {Zoccali},
  {S{\"o}kmen}, {Hidalgo}, \& {Minniti}}]{Surot2020}
{Surot}, F., {Valenti}, E., {Gonzalez}, O.~A., {et~al.} 2020, \aap, 644, A140

\bibitem[{{Vasiliev} \& {Baumgardt}(2021)}]{2021MNRAS.505.5978V}
{Vasiliev}, E. \& {Baumgardt}, H. 2021, \mnras, 505, 5978

\bibitem[{{Virtanen} {et~al.}(2020){Virtanen}, {Gommers}, {Oliphant},
  {Haberland}, {Reddy}, {Cournapeau}, {Burovski}, {Peterson}, {Weckesser},
  {Bright}, {van der Walt}, {Brett}, {Wilson}, {Jarrod Millman}, {Mayorov},
  {Nelson}, {Jones}, {Kern}, {Larson}, {Carey}, {Polat}, {Feng}, {Moore}, {Vand
  erPlas}, {Laxalde}, {Perktold}, {Cimrman}, {Henriksen}, {Quintero}, {Harris},
  {Archibald}, {Ribeiro}, {Pedregosa}, {van Mulbregt}, \&
  {Contributors}}]{Virtanen_2020}
{Virtanen}, P., {Gommers}, R., {Oliphant}, T.~E., {et~al.} 2020, Nature
  Methods, 17, 261

\bibitem[{{Wegg} \& {Gerhard}(2013)}]{wegg2013}
{Wegg}, C. \& {Gerhard}, O. 2013, \mnras, 435, 1874

\bibitem[{{Wylie} {et~al.}(2022){Wylie}, {Clarke}, \& {Gerhard}}]{Wylie2022}
{Wylie}, S.~M., {Clarke}, J.~P., \& {Gerhard}, O.~E. 2022, \aap, 659, A80

\bibitem[{{Zhang} \& {Kainulainen}(2022)}]{zhang2022}
{Zhang}, M. \& {Kainulainen}, J. 2022, \mnras, 517, 5180

\bibitem[{{Zoccali}(2019)}]{Zoccali2019}
{Zoccali}, M. 2019, Boletin de la Asociacion Argentina de Astronomia La Plata
  Argentina, 61, 137

\bibitem[{{Zoccali} {et~al.}(2003){Zoccali}, {Renzini}, {Ortolani}, {Greggio},
  {Saviane}, {Cassisi}, {Rejkuba}, {Barbuy}, {Rich}, \& {Bica}}]{Zoccali03}
{Zoccali}, M., {Renzini}, A., {Ortolani}, S., {et~al.} 2003, \aap, 399, 931

\bibitem[{{Zoccali} \& {Valenti}(2016)}]{Zoccali+Valenti2016}
{Zoccali}, M. \& {Valenti}, E. 2016, \pasa, 33, e025

\end{thebibliography}

\begin{appendix}

\section{Color-magnitude diagrams of the studied fields}
\label{sec:appendixBTP}

In this section, we show the HST and \textit{Gaia} DR3 CMDs for the remaining fields analysed in this study: BTP-Stanek window, BTP-Ogle29 window, $\omega\,$Cen F1, $\omega\,$Cen F2, $\omega\,$Cen F3 and NGC 6652. 
The black points are all the sources in a given field after the quality cuts described in Sect. 2, while the orange points are the final cross-matched sources used for the study. The figures are the equivalent to Fig. \ref{fig:cmds_sweeps}, where the left and right panels correspond to the CMD in HST and \textit{Gaia} DR3 filters, respectively. In addition, we also show the CMD for Baade's window BTP catalogue in comparison with \textit{Gaia}. This field was not used for the analysis due to too few cross-matched sources.

\begin{figure}
\centering
\includegraphics[width=0.5\textwidth]{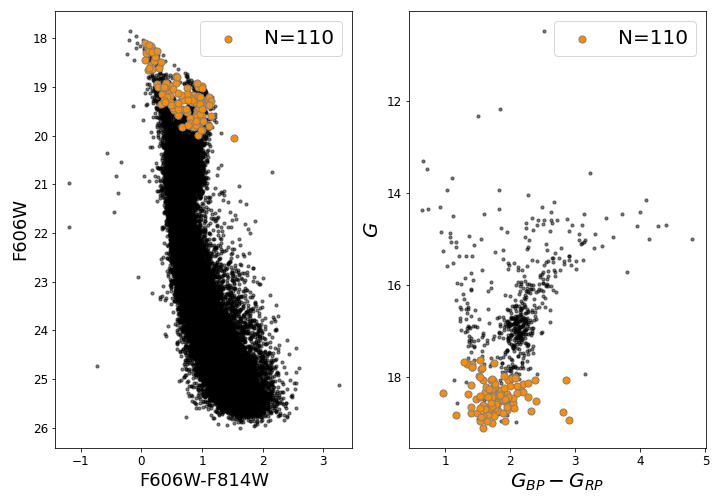}
\caption{Colour Magnitude Diagrams of \textit{Gaia} and HST in the Stanek's window. The orange points correspond to the location of the cross-match after the quality cuts and 3$\sigma$ clipping.}
\label{fig:cmds_stanek}
\end{figure}

\begin{figure}
\centering
\includegraphics[width=0.5\textwidth]{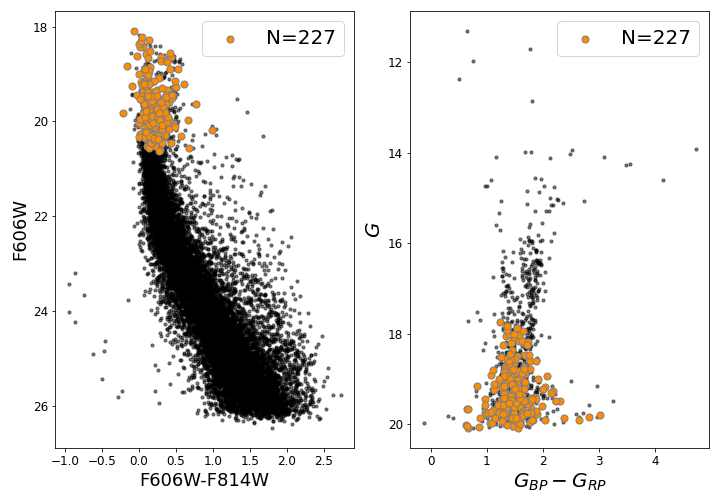}
\caption{Colour Magnitude Diagrams of \textit{Gaia} and HST in the Ogle29 window. The orange points correspond to the location of the cross-match after the quality cuts and 3$\sigma$ clipping.}
\label{fig:cmds_ogle29}
\end{figure}

\begin{figure}
\centering
\includegraphics[width=0.5\textwidth]{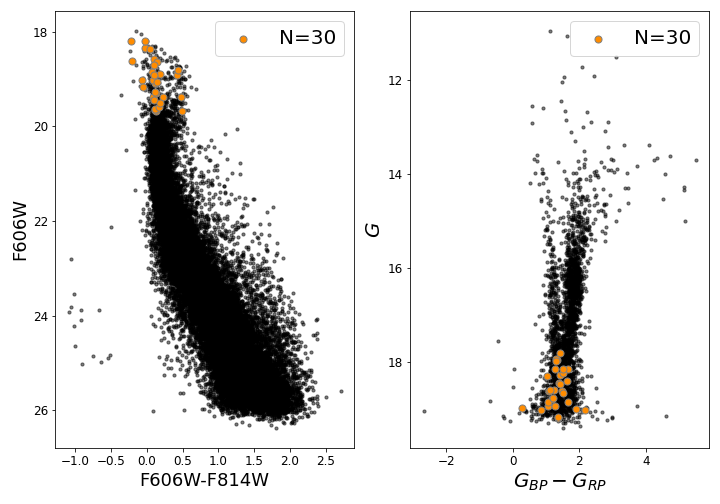}
\caption{Colour Magnitude Diagrams of \textit{Gaia} and HST in the Baade's window. The orange points correspond to the location of the cross-match after the quality cuts and 3$\sigma$ clipping.}
\label{fig:cmds_baade}
\end{figure}

\begin{figure}
\centering
\includegraphics[width=0.5\textwidth]{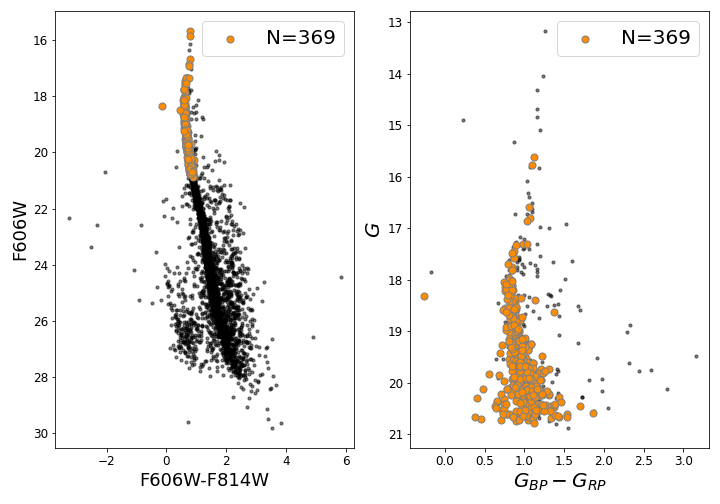}%
\caption{Colour Magnitude Diagrams of \textit{Gaia} and HST in the $\omega\,$Cen F1  field. The orange points correspond to the location of the cross-match after the quality cuts and 3$\sigma$ clipping.}
\label{fig:}
\end{figure}

\begin{figure}
\centering
\includegraphics[width=0.5\textwidth]{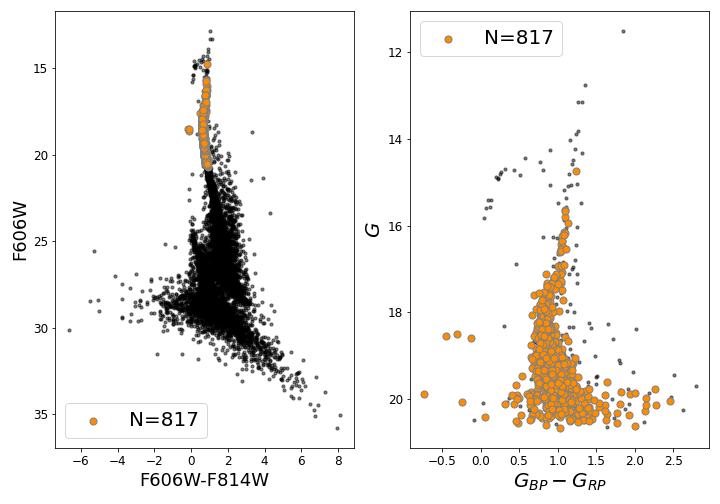}%
\caption{Colour Magnitude Diagrams of \textit{Gaia} and HST in the $\omega\,$Cen F2 field. The orange points correspond to the location of the cross-match after the quality cuts and 3$\sigma$ clipping.}
\label{fig:}
\end{figure}

\begin{figure}
\centering
\includegraphics[width=0.5\textwidth]{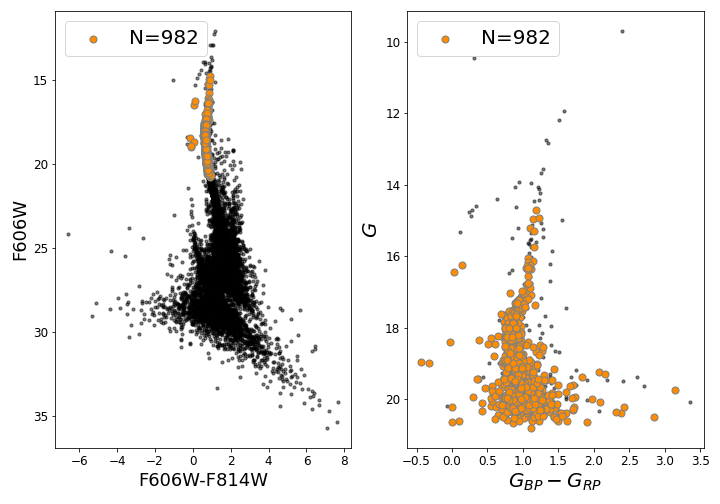}%
\caption{Colour Magnitude Diagrams of \textit{Gaia} and HST in the $\omega\,$Cen F3 field. The orange points correspond to the location of the cross-match after the quality cuts and 3$\sigma$ clipping.}
\label{fig:}
\end{figure}

\begin{figure}
\centering
\includegraphics[width=0.5\textwidth]{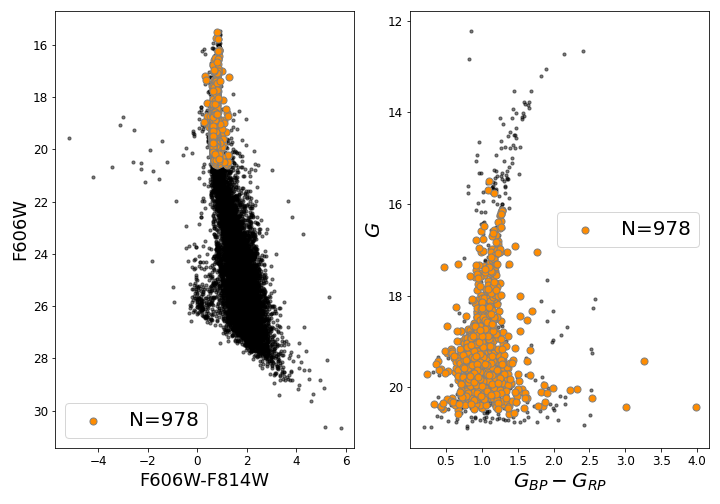}%
\caption{Colour Magnitude Diagrams of \textit{Gaia} and HST in the globular cluster NGC\,6652. The orange points correspond to the location of the cross-match after the quality cuts and 3$\sigma$ clipping.}
\label{fig:}
\end{figure}

\section{Color-magnitude diagrams of the studied fields with VIRAC2}
\label{sec:appendix_cdm_VIRAC2}

In this section, we show the HST and VIRAC2 CMDs for the  BTP SWEEPS, Stanek and Ogle29 fields. 
The black points are all the sources in a given field after the quality cuts described in Sect. 2, while the orange points are the final cross-matched sources used for the study. The figures are the equivalent to Fig. \ref{fig:cmds_sweeps}, where the left and right panels correspond to the HST and VIRAC2 CMDs, respectively.

\begin{figure}
\centering
\includegraphics[width=0.5\textwidth]{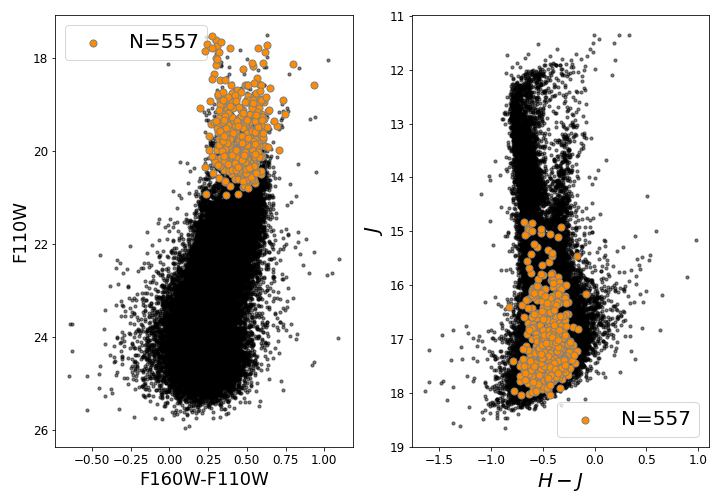}%
\caption{Colour Magnitude Diagrams of VIRAC2 and HST in the SWEEPS window. The orange points correspond to the location of the cross-match after the quality cuts and 3$\sigma$ clipping.}
\label{fig:cmd_virac_sweeps}
\end{figure}

\begin{figure}
\centering
\includegraphics[width=0.5\textwidth]{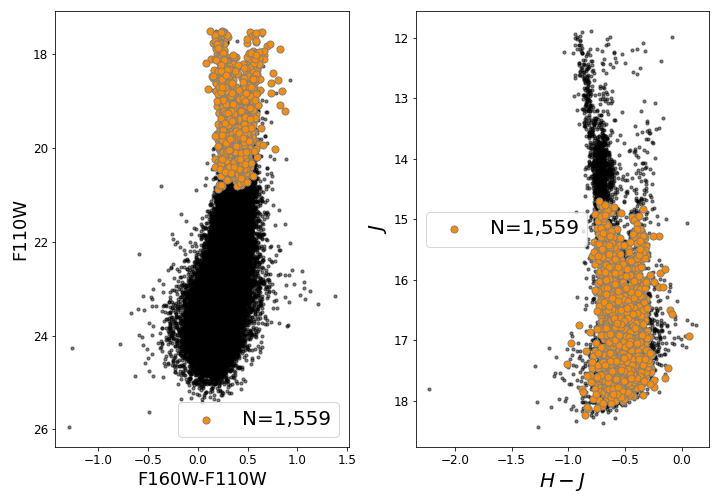}%
\caption{Colour Magnitude Diagrams of VIRAC2 and HST in the Stanek window. The orange points correspond to the location of the cross-match after the quality cuts and 3$\sigma$ clipping.}
\label{fig:cmd_virac_stanek}
\end{figure}

\begin{figure}
\centering
\includegraphics[width=0.5\textwidth]{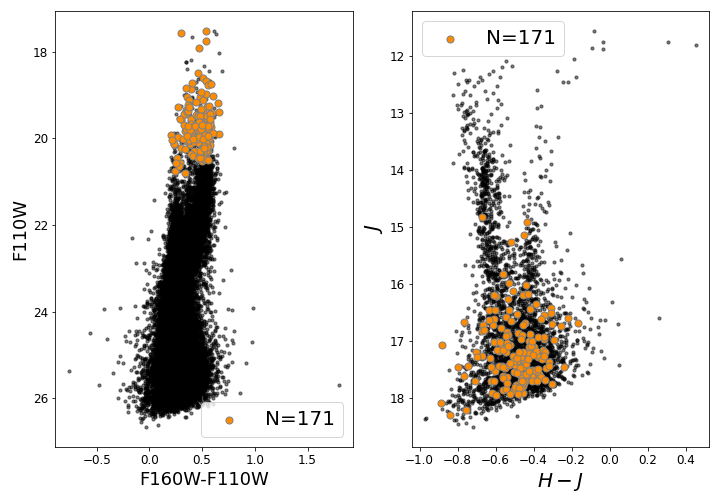}%
\caption{Colour Magnitude Diagrams of VIRAC2 and HST in the Ogle29 window. The orange points correspond to the location of the cross-match after the quality cuts and 3$\sigma$ clipping.}
\label{fig:cmd_virac_ogle}
\end{figure}

\section{Proper motion comparison between \textit{Gaia} DR3 and HST.}
\label{sec: pm comp}

\begin{figure}
\centering
\includegraphics[width=0.5\textwidth]{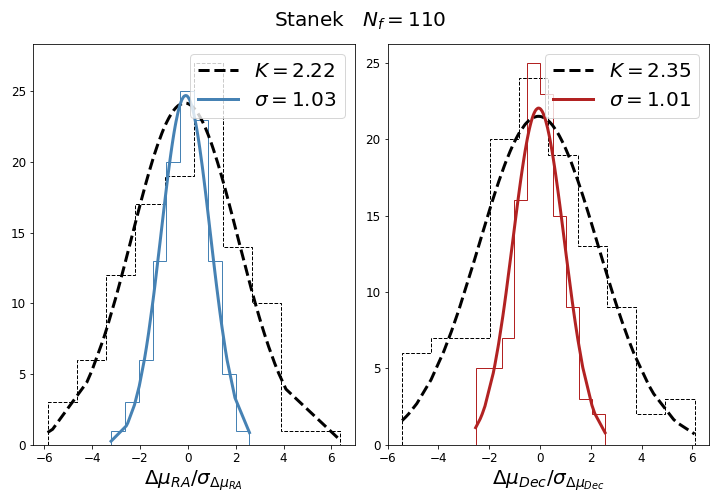}
\caption{Same as Fig. \ref{fig:pm_comparison}, but for Stanek window.}
\label{fig:pm_comparison_st}
\end{figure}

\begin{figure}
\centering
\includegraphics[width=0.5\textwidth]{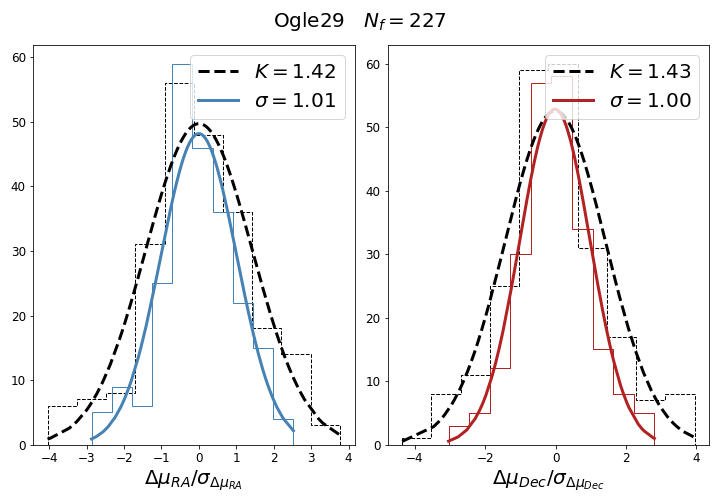}
\caption{Same as Fig. \ref{fig:pm_comparison}, but for Ogle29 window.}
\label{fig:pm_comparison_o29}
\end{figure}

\begin{figure}
\centering
\includegraphics[width=0.5\textwidth]{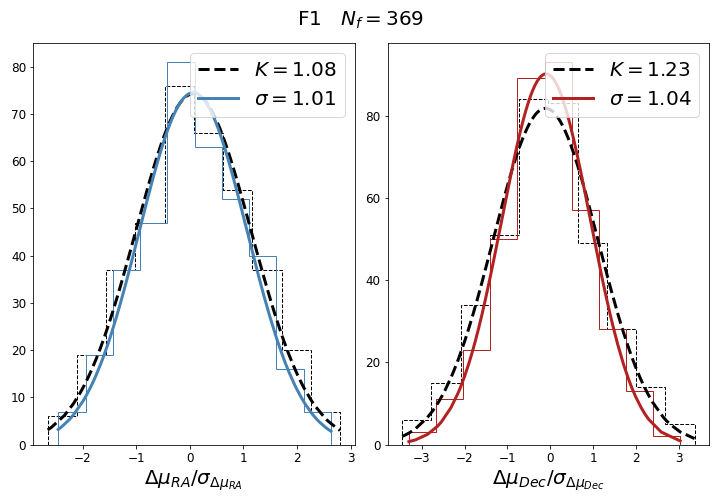}
\caption{Same as Fig. \ref{fig:pm_comparison}, but for $\omega\,$Cen F1 field.}
\label{fig:pm_comparison_f1}
\end{figure}

\begin{figure}
\centering
\includegraphics[width=0.5\textwidth]{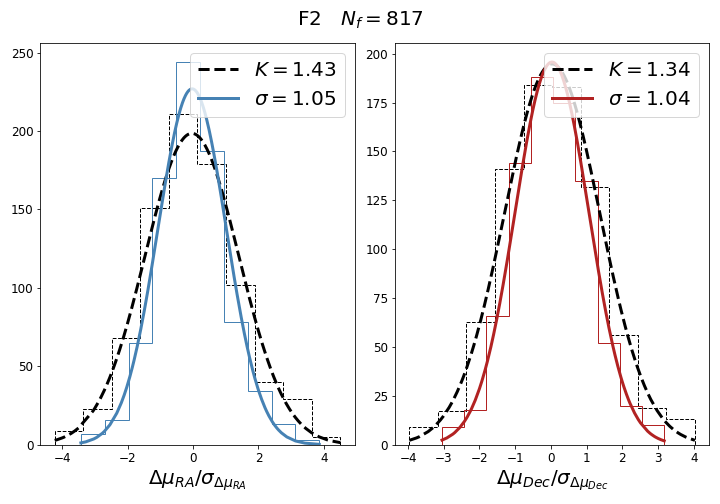}
\caption{Same as Fig. \ref{fig:pm_comparison}, but for $\omega\,$Cen F2 field.}
\label{fig:pm_comparison_f2}
\end{figure}

\begin{figure}
\centering
\includegraphics[width=0.5\textwidth]{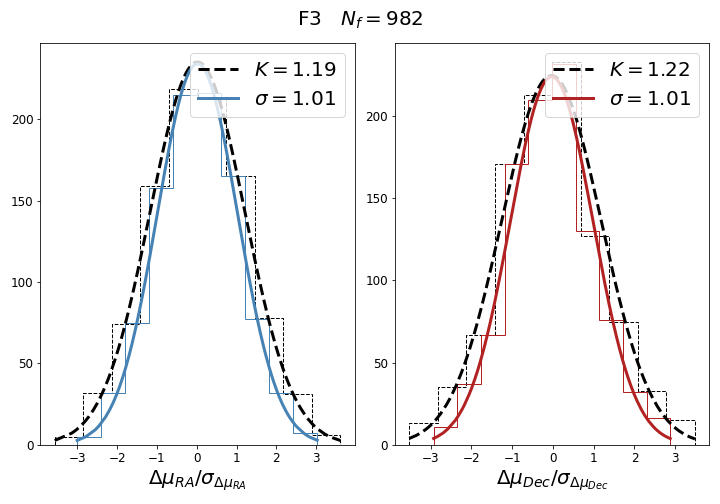}
\caption{Same as Fig. \ref{fig:pm_comparison}, but for $\omega\,$Cen F3 field.}
\label{fig:pm_comparison_f3}
\end{figure}

\begin{figure}
\centering
\includegraphics[width=0.5\textwidth]{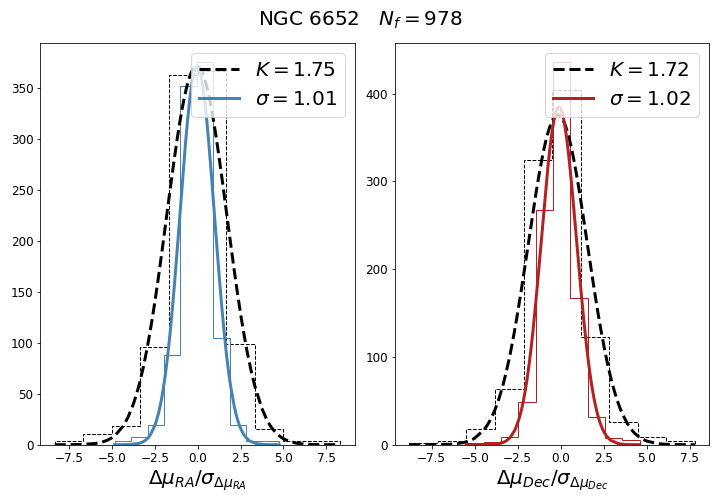}
\caption{Same as Fig. \ref{fig:pm_comparison}, but for the globular cluster NGC 6652.}
\label{fig:pm_comparison_n6652}
\end{figure}

In this section, we show the uncertainty normalised proper motion differences  ($\Delta \mu/ \sigma \mu$) between \textit{Gaia} DR3 and HST for the fields: BTP-Stanek window, BTP-Ogle29 window, $\omega\,$Cen F1, $\omega\,$Cen F2, $\omega\,$Cen F3 and NGC 6652. The Figures \ref{fig:pm_comparison_st}, \ref{fig:pm_comparison_o29}, \ref{fig:pm_comparison_f1}, \ref{fig:pm_comparison_f2}, \ref{fig:pm_comparison_f3}, and \ref{fig:pm_comparison_n6652} are the equivalent to Fig. \ref{fig:pm_comparison}, where the left and right panels correspond to $\Delta \mu/ \sigma \mu$ in RA and Dec, respectively.

\section{Variation of $r$ in the BTP}
\label{sec:appendixBTP_rvar}

\begin{figure}
    \centering
    \includegraphics[width=0.4\textwidth]{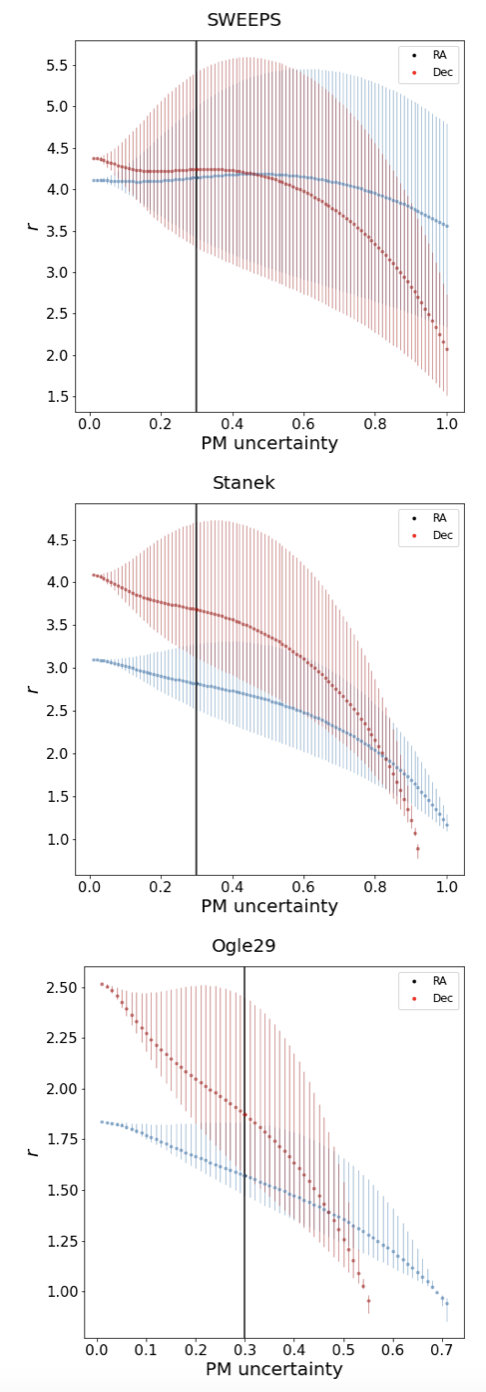}
    \caption{Variation of $r$ in the BTP fields if different upper limits of PM uncertainties are adopted. The PM uncertainty is given in $\mathrm{mas\,yr^{-1}}$. The points are the median $r$ at a given $\sigma_{\mu}$, and the vertical lines mark the range between the 16$^{th}$ and 84$^{th}$ percentiles.}
    \label{fig:rvar}
\end{figure}

Here, we explore how the inflation factor varies if we adopt an upper limit in the BTP PM uncertainties different than $0.3~\mathrm{mas\,yr^{-1}}$, within the range $0.3-1 ~\mathrm{mas\,yr^{-1}}$. This is to account for HST systematics that we cannot quantify in the data. 
For the Stanek and Ogle29 datasets, the inflation factor is reduced up to 20\% if we assume a BTP PM uncertainty of $0.6~\mathrm{mas\,yr^{-1}}$, and up to 50\% if we assume a BTP PM uncertainty of $1~\mathrm{ mas\,yr^{-1}}$; however, for the SWEEPS dataset, the inflation factor is reduced only by 25\% assuming a PM uncertainty of $1~\mathrm{mas\,yr^{-1}}$. The results are summarised in Figure \ref{fig:rvar}.

\section{Comparison of other fields with the extended SWEEPS catalogue}
\label{sec:appendixSWEEPSext}

\begin{figure*}[]
\centering
\includegraphics[width=0.9\textwidth]{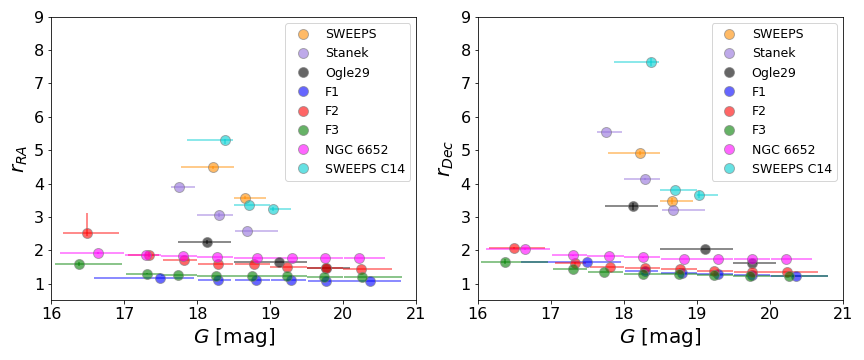}
\caption{Inflation factor $r$ dependence on $G$-band magnitude of the studied HST fields. Same as Fig. \ref{fig:r_G_all_btsp}, but with the addition of the SWEEPS extended catalogue (SWEEPS C14 in the label). The points represent the medians of the magnitude bins, and for $r$, the error bars are of the 16$^{th}$  and 84$^{th}$ percentiles. The bars in $G$ indicate the magnitude distribution at a given bin, where the marker is the median.}
\label{fig:r_G_all_btsp_swfull}
\end{figure*}

In this section, we extend the discussion of Sect.\ref{sec:sweeps_full}. The extended SWEEPS catalogue \citep{Calamida_SWEEPS2014} is based on observations of the BTP-SWEEPS window with additional epochs that extended the time baseline leading to more accurate measurements with individual uncertainties well characterised. This dataset provides an opportunity to internally validate the BTP-SWEEPS data. 
The median inflation factor of the SWEEPS extended catalogue computed in different magnitude bins is labelled as SWEEPS C14 in Fig. \ref{fig:r_G_all_btsp_swfull}. The median inflation factor is consistent with that of the BTP-SWEEPS window adopting a PM uncertainty upper limit of $0.3~\mathrm{mas\,yr^{-1}}$. The exception is found for the brightest bin in the Dec component ($G<18.5$), where some sources may have poor quality, but we cannot filter them due to the lack of quality flags in the extended SWEEPS catalogue.

\end{appendix}

\end{document}